\documentclass[journal]{IEEEtran}
\usepackage[noadjust]{cite}

\ifCLASSINFOpdf
\else
\fi
\usepackage{amsmath} 
\usepackage{graphicx}
\usepackage{booktabs} 
\usepackage{amsmath}
\usepackage{mathtools}
\usepackage{algpseudocode}
\usepackage{algorithm}
\usepackage{algorithmicx}
\usepackage{textcomp}
\usepackage{microtype}
\usepackage{float}
\usepackage{adjustbox}
\usepackage{booktabs,makecell,tabularx}
\usepackage{url}
\usepackage{booktabs}
\usepackage{subfigure}

\newcolumntype{C}[1]{>{\centering\arraybackslash}p{#1}}
\newcolumntype{L}{>{\raggedright\arraybackslash}X}
\usepackage{siunitx}
\usepackage[utf8]{inputenc}

\usepackage{etoolbox}
\usepackage{xparse}
\makeatletter

\usepackage[utf8]{inputenc}
\usepackage[T1]{fontenc}
\usepackage{regexpatch,fancyvrb,xparse}
\usepackage{nomencl}
\usepackage{enumitem}
\usepackage{amsmath,amssymb}
\makenomenclature

\usepackage{etoolbox}
\renewcommand\nomgroup[1]{%
  \item[\Large\bfseries
  \ifstrequal{#1}{A}{Abbreviations}{%
  \ifstrequal{#1}{N}{Nomenclature}{}}%
]\vspace{10pt}} 
\usepackage[utf8]{inputenc}
\usepackage[english]{babel}

\usepackage{amsthm}

\begin{document}
%
\title{Eco-Coasting Strategies Using Road Grade Preview: Evaluation and Online Implementation Based on Mixed Integer Model Predictive Control}
%
%
%

\author{Yongjun Yan, Nan Li, Jinlong Hong, Bingzhao Gao, Hong Chen, Jing Sun, and Ziyou Song
\thanks{Y. Yan is with the State Key Laboratory of Automotive Simulation and Control, Jilin University, Changchun 130022, China. This work was done when he is a visiting Ph.D. student in the Department of Naval Architecture and Marine Engineering, University of Michigan, Ann Arbor, MI 48109 USA (e-mail: yanyj18@mails.jlu.edu.cn).}
\thanks{N. Li is with the Department of Aerospace Engineering, University of Michigan, Ann Arbor, MI 48109 USA (e-mail: nanli@umich.edu)).}
\thanks{J. Hong, and B. Gao are with the Clean Energy Automotive Engineering Center, Tongji University, Shanghai, 201804, China and also with the State Key Laboratory of Automotive Simulation and Control, Jilin University, Changchun 130022, China (e-mail: hongjl@tongji.edu.cn; gaobz@jlu.edu.cn).}
\thanks{H. Chen is with the College of Electronic and Information Engineering, Tongji University, Shanghai 201804, China (e-mail: chenhong2019@tongji.edu.cn).}
\thanks{J. Sun is with the Department of Naval Architecture and Marine Engineering, University of Michigan, Ann Arbor, MI 48109 USA (e-mail: jingsun@umich.edu).}
\thanks{Z. Song is with the Department of Mechanical Engineering, National University of Singapore, Singapore 117575, Singapore and also with the Department of Naval Architecture and Marine Engineering, University of Michigan, Ann Arbor, MI 48109 USA (e-mail: Ziyou@nus.edu.sg).}}

%
%

\markboth{Journal of \LaTeX\ Class Files,~Vol.~14, No.~8, August~2015}%
{Shell \MakeLowercase{\textit{et al.}}: Bare Demo of IEEEtran.cls for IEEE Journals}
%



\maketitle

\begin{abstract}
Coasting has been widely used in the eco-driving guidelines to reduce fuel consumption by profiting from kinetic energy. However, the comprehensive comparison between different coasting strategies and online performance of the eco-coasting strategy using road grade preview are still unclear because of the oversimplification and the integer variable in the optimal control problems. Herein, two different coasting strategies (fuel cut-off and engine start/stop) are proposed to reveal the potential benefit of eco-coasting using the road grade preview. Engine drag torque and energy cost used for engine restart are considered in the modeling to give a fair evaluation of the offline and online performance. The offline performance of these two coasting methods is evaluated through dynamic programming (DP) under various driving scenarios with different slope profiles. Offline simulation shows that the engine start/stop method outperforms the fuel cut-off method in terms of fuel consumption and travel time by getting rid of the engine drag torque. Then, online performance of these two coasting methods is evaluated using Mixed Integer Model Predictive Control (MIMPC). A novel operational constraint on the minimum off steps is added in the MIMPC formulation to avoid frequent switch of the integer variables which represent the fuel cut-off and the engine start/stop mechanism. Simulation results show that, for both fuel cut-off and engine start/stop coasting methods, the MPC controller reduces fuel consumption to a level comparable to DP without sacrificing the travel time.
\end{abstract}


%
\IEEEpeerreviewmaketitle

\vspace{-0.2cm}
\section*{Nomenclature}
\begin{tabular*}{0.9\textwidth}{ l p{11cm}}
$A$ & Frontal projected area of vehicle, [$m^2$].\\
$C$ & Vector of constraints, [$-$].\\
$C_{d}$ & Coefficient of drag, [$-$].\\
$d_{\text{min}}$ & Minimum off steps, [$-$].\\
$d$ & Engine start/stop signal, [$-$].\\
$F_{t}$ & Traction force, [$N$].\\
$F_{r}$ & Resistance force, [$N$].\\
$f_{r}$ & Coefficient of rolling resistance, [$-$].\\
$g$ & Gravitational constant, [$-$].\\
$I_{\text{engine}}$ & Engine inertia, [$kg \, m^2$].\\
$I_{\text{final}}$ & Gear ratio of final reduction drive, [$-$].\\
$I_{g}$ & Gear ratio of gearbox, [$-$].\\
$M_{f}$ & Fuel consumption model in the distance \\
                     &  domain, [$-$].\\
$m_\text{eff}$ & Effective mass of the vehicle, [$kg$].\\
$\dot{m}_{{f}}$ & Instantaneous fuel consumption rate, [$g/s$].\\
$N_{h}$ & Prediction horizon, [$-$].\\
$r$ & Wheel radius, [$m$].\\
$s$ & Space coordinate, [$m$].\\
$T_{b}$ & Brake torque, [$Nm$].\\
$T_{e}$ & Engine torque, [$Nm$].\\
$T_{\text{final,1}}$ & Final driveshaft torque of baseline \\
                     &  method, [$Nm$].\\
$T_{\text{final,2}}$ & Final driveshaft torque of fuel cut-off \\
                     &  method, [$Nm$].\\
$T_{\text{final,3}}$ & Final driveshaft torque of engine\\
                     &  start/stop method, [$Nm$].\\
$v$ & Vehicle longitudinal velocity, [$m/s$].\\
$\bar{v}$ & Partial velocity variation, [$m/s$].\\
$z$ & fuel cut-off signal, [$-$].\\
$z_{\text{current}}$ & Optimal binary variables among the current\\
                     &  horizon, [$-$].\\
$z_{\text{history}}$ & History binary variables, [$-$].\\
$\alpha_{1 \sim 4}$ & Coefficients of fuel consumption model, [$-$].\\
$\beta$ & Weight of weighted sum method, [$-$].\\
$\Delta E_{\text {engine }}$ & Energy required for cranking the engine, [$J$].\\
$\Delta{s}$ & Discretization distance, [$-$].\\
$\Delta{v}$ & Velocity reduction, [$m/s$].\\
$\delta E_{v}$ & Partial inertia energy, [$J$].\\
$\delta $ & Extended 0/1 variables, [$-$].\\
$\varepsilon$ & Tolerance constant, [$-$].\\
$\eta$ & Gearbox efficiency, [$-$].\\
$\phi$ & Slope data, [$^\circ$].\\
$\rho$ & Air density, [$[kg/m^3]$].\\
$\omega_{e}$ & Rotational velocity of the engine, [$rpm$].\\
$\omega_{w}$ & Rotational velocity of wheel, [$rpm$].\\

\end{tabular*}
\vspace{-0.30cm}
\section*{Acronyms}
\begin{tabular*}{0.9\textwidth}{ l p{11cm}}
\textit{$BSFC$} & Brake Specific Fuel Consumption.\\
\textit{$DP$} & Dynamic Programming.\\
\textit{$FCO$}  & Fuel Cut-Off.\\
\textit{$MPC$} & Model Predictive Control.\\
\textit{$MIMPC$} & Mixed Integer Model Predictive Control.\\
\textit{$MIOCP$} & Mixed Integer Optimal Control Problems.\\
\textit{$SUV$} & Sports Utility Vehicle.\\
\textit{$TCU$} & Transmission Control Unit.\\
\end{tabular*}

\section{Introduction}
%
%
%
%
\IEEEPARstart{F}{or} on-road vehicles, the key to further improve fuel economy with eco-driving is comprehensive incorporation and full integration of the powertrain system with the driving environment represented by traffic conditions and digital maps. Eco-coasting, as a category of eco-driving, refers to the strategies to roll the vehicle with kinetic energy without traction force~\cite{varnhagen2017electronic}. For a conventional car with a gasoline engine and automatic transmission, there are three methods to coast during the deceleration phases: setting the gear in the neutral position and turning the engine to idle~\cite{salgueiredo2017experimental, li2015effect}; shutting fuel injection off when no torque is requested~\cite{balluchi1997cut, katsumata2008development, koch2014criteria}; manipulating the lock-up clutch to disconnect the engine from powertrain when the engine is off~\cite{lee2009vehicle,bishop2007engine,griefnow2017next,griefnow2019real,sohn2020analysis}.

The majority of studies on eco-coasting have focused on the fuel economy benefit evaluation of different coasting strategies with backward vehicle simulation approaches ~\cite{salgueiredo2017experimental, choi2013optimal, mueller2011next}, where the required engine load is calculated from the pre-defined vehicle speed profile. However, it does not fully explore the potential of eco-coasting as the generated coasting action only relied on current vehicle state and road information. Previous investigation has proven that the look-ahead control can achieve better performance when the slope information on the road ahead is available~\cite{hellstrom2010design}. To the author’s best knowledge, no research has been reported on the comparison of the fuel saving potential among different coasting methods under the cruising condition with road grade preview, let alone the online implementation of these coasting strategies. 

Model predictive control (MPC) is a promising method for the online implementation of eco-coasting strategies. However, the discrete nature of the engine start/stop and fuel cut-off coasting methods render the mixed-integer optimal control problems (MIOCPs). Similar to the gear shift optimization~\cite{ngo2012gear,li2017hybrid} or engine start/stop of hybrid vehicles~\cite{ngo2012gear,yan2012hybrid} in the online energy management field. There are two open problems for the online implementation of the eco-coasting strategies with MPC controller: (1) frequent switch of the integer control variables; and (2) computational burden. To mitigate the frequent switch of the integer signal, an additional penalty of the switching frequency~\cite{ngo2012gear} and inequality constraints implying that the limitation of total switch numbers of the integer variables is imposed in the optimization problems~\cite{esmaeili2018optimal}. These formulations yield a reduction in the switching frequency for a finite horizon open-loop optimal control problem. However, neither the additional penalty nor the inequality constraints is effective for MPC when the first element of the control sequence will change at different sampling instant if the solution is numerically sensitive to different driving conditions. In~\cite{parisio2016stochastic,mayer2015management}, mixed-integer linear inequalities are applied in the MPC formulation, modeling the minimum amount of time for which a on/off control variable must be kept on/off.

As to the computational complexity, generally, there are two approaches to solve the MINLP: the first one is to solve the MINLP directly with branch \& bound method or the related solvers that require excessive computational resource~\cite{belotti2013mixed}; the second one is the heuristic method which reformulates the original MINLP into NLP by relaxing the integer variables as continuous ones, then round the feasible solution to the nearest integer. This relax-and-round method is computation efficient but can not guarantee the optimality of the final solution~\cite{takapoui2020simple,yu2020mixed}. In~\cite{deng2020flexible,elbert2014engine,hadj2016convex,zhang2021hierarchical}, engine start/stop signal is calculated in the upper layer with Dynamic Programming (DP). With a heuristic engine start/stop strategies, the original mixed-integer hybrid powertrain optimization problem can be converted into a simple convex optimization problem. Therefore, the problems referring to the frequent switch and computational burden are mitigated, but the robustness issues arising from the heuristic feature is challenging.

In this paper, instead of assuming the drag torque is zero in the fuel cut-off method~\cite{han2019fundamentals} or ignoring the extra cost to restart the engine~\cite{jia2020energy} for energy management research, two coasting models are proposed to capture the fuel cut-off and engine start/stop mechanisms. Comprehensive performance concerning the travel time and fuel consumption of these two coasting methods are evaluated by solving a multi-objective optimization problem with DP using different slope profiles which are known as a priori. Finally, the online implementation of the aforementioned coasting methods is investigated using an MPC framework.

The contributions of this paper are as follows:\\
\hspace*{1em}1)	We quantify the potential benefits of these two coasting methods based on results derived by DP under different slope profiles from digital maps. \\
\hspace*{1em}2)	We assess the online implementation for these two coasting methods with the MIMPC scheme compared with the PI controller. \\
\hspace*{1em}3) A novel MIMPC formulation with inequality constraints applied to the integer variables is used to mitigate the frequent switch of the coasting signal.\\
\hspace*{1em}4) Through simulations, we demonstrate that the computation burden can be reduced dramatically using offline calculated coasting signal when the slope profile and reference speed are fixed in advance.\\

The remainder of this paper is organized as follows: Section \uppercase\expandafter{\romannumeral2} discusses different vehicle coasting models; Section \uppercase\expandafter{\romannumeral3} describes the offline evaluation of the proposed coasting strategies; Section \uppercase\expandafter{\romannumeral4} presents the online implementation of two proposed coasting methods with MIMPC controller; Finally, conclusions and future work are concluded in Section \uppercase\expandafter{\romannumeral5}.

\subsection{Vehicle Model}

Consider the longitudinal vehicle model described as
\begin{equation}\begin{aligned}
\dot{s} &=v, \\
\dot{v} &=\frac{1}{m_{\text{eff}}}\left[F_{t}-F_{r}(s, v)\right],
\end{aligned}\end{equation}
where $s$ is the distance traveled and $v$ is the velocity, $\dot{s}$ and $\dot{v}$ are the derivative with respect to time $t$, $m_\text{eff}$ is the total mass of the vehicle, $F_{t}$ is the traction force and $F_{r}$ is the total resistance force.

\begin{table}[htbp]
\centering
\caption{Parameters of Vehicle}
\label{Parameters of Vehicle}
\small
\begin{tabular}{c c c}
\toprule
Symbol  & Description  & Value\\\hline
$m_\text{eff}$   &  Effective mass of the vehicle         & 1870 $kg$   \\
$C_{d}$     &  Coefficient  of  drag                 & 0.373 \\
$A$         &  Frontal projected area of vehicle     & 2.58 $m^{2}$ \\
$\rho$      &  Air density                           & 1.205 $kg / m^{3}$ \\
$f_{r}$     &  Rolling resistance coefficient        & 0.011 \\
$g$         &  Gravitational  constant               & 9.8 $m^{3} kg^{-1} s^{-2}$\\
$\eta$      &  Gearbox efficiency                    & 0.94 \\
$r_{w}$         &  Wheel radius                          & 0.364 $m$ \\
$I_{g}$     &  Gear ratio of gearbox                 & 0.672 \\
$I_{\text{final}}$ &  Final driver ratio   & 4.103 \\
$T_{\text{drag}}$  &  Engine drag torque                    & 30 $Nm$\\
$I_{\text{engine}}$&  Engine inertia                        & 0.15 $kg \, m2$\\
$\alpha_{1}$&  Coefficient  & 0.2159 \\
$\alpha_{2}$&  Coefficient  & 0.005676 \\
$\alpha_{3}$&  Coefficient  & 0.0004349 \\
$\alpha_{4}$&  Coefficient  & 8.899E-07 \\
\bottomrule
\end{tabular}
\end{table}

The total resistance force, $F_{r}$, consists of the aerodynamic resistance, the grade resistance, and the rolling resistance, as calculated by 
\begin{equation}
\label{dynamic time}
F_{r}(s, v)=m_{\text{eff}} g\left(\sin (\phi(s))+f_{r}\cos (\phi(s))\right)+\frac{C_{d} \rho A}{2} v^{2},
\end{equation}
where $g$ is the gravitational constant, $\phi(s)$ is the slope data in the distance domain, $f_{r}$ is the coefficient of rolling resistance, $C_{d}$ is the drag coefficient, $\rho$ is the air density, and $A$ is the frontal projected area of the vehicle. Assuming no slip on the wheel, the wheel rotational speed $\omega_{w}$ and traction force $F_{t}$ can be calculated as:
\begin{equation}\begin{aligned}
\omega_{w} &=\frac{v}{r_{w}}, \\
F_{t} &=\frac{1}{r_{w}} T_{\text{final}},
\end{aligned}\end{equation}
where $T_{\text {final }} \in\left\{T_{\text {final }, 1}, T_{\text {final }, 2}, T_{\text {final, } 3}\right\}$ is the final driveshaft torque for baseline, fuel cut-off, and engine start/stop methods respectively, and $r_{w}$ is the wheel radius.

The vehicle dynamic equations in (\ref{dynamic time}) are defined in the time coordination. However, the slope information represented in the digital map is location-specific, and is more naturally correlated to spatial information. To that end, we adopt the spatial-based model, as shown below
\begin{equation}
\frac{\text{d} v}{\text{d} t}=\frac{\text{d} v}{\text{d} s} \frac{\text{d} s}{d t}=\frac{\text{d} v}{\text{d} s} v.
\end{equation}

The model in the spatial domain can be given as
\begin{equation}
\label{Dynamic model}
\begin{split}
\frac{\text{d} v}{\text{d} s} = \frac{1}{m_\text{eff} v}\left[F_{t}-F_{r}(s, v)\right] .
\end{split}
\end{equation}

By applying Euler forward method and sampling in the $s$-domain, the vehicle dynamic model is discretized as
\begin{equation}
\label{Discrete dynamic vehicle-level model}
\begin{split}
v(k+1) & =g\left(v(k), T_{e}(k), T_{b}(k)\right)\\
& =v(k)+\frac{1}{m_\text{eff} v(k)}\left[F_{t}-F_{r}(k d s, v(k))\right] \Delta s,
\end{split}
\end{equation}
where $\Delta s$ is the discretization distance in each step. 

\subsection{Powertrain Model}
A quasi-static engine model is considered to quantify fuel consumption. A polynomial fuel consumption model is fitted using the measurements on an engine dynamometer provided by the manufacturer. The idle fuel mass flow rate, which is absent from the engine dynamometer measurements, is measured with fuel meter during on-road tests. The following model is constructed considering the computational complexity and fuel cut-off mechanism: 
\begin{equation}
\dot{m}_{f}=\alpha_{1}+\alpha_{2} \omega_{e} T_{e}+\alpha_{3} \omega_{e}^{2} T_{e}+\alpha_{4} \omega_{e} T_{e}^{2}
\end{equation}
where $\dot{m}_{f}$ $[g/s]$ is the instantaneous fuel consumption rate, and $\alpha_{1 \sim 4}$ are the constant coefficients as shown in Table~\ref{Parameters of Vehicle}.

Note that:
\begin{itemize}
    \item[$\blacktriangleright$] A constant term is included in the polynomial equation to represent the idle fuel consumption;
    \item[$\blacktriangleright$] A 3rd-order polynomial is used to avoid over-fitting and to reduce computation burden;
    \item[$\blacktriangleright$] Engine torque impacts every term besides the constant one. 
\end{itemize}

The reformulated fuel consumption in the distance domain is represented as 
\begin{equation}
M_{f}(k)=\frac{\dot{m}_{f}\left(T_{e}(k)\right)}{v(k)} \Delta s
\end{equation}
where $M_{f}(k)$ is the fuel consumption over the discretization step.

In this study, three different coasting methods are considered and compared.

\subsubsection{Baseline Model}
When the vehicle cruises on the highway, the transmission is set at the highest gear which corresponds to the high efficient area in the Brake Specific Fuel Consumption (BSFC) map. The lock-up clutch keeps engaged. Therefore, the drivetrain model is given by
\begin{equation}
\label{dynamic equation of baseline model}
\begin{aligned}
T_{\text{final},1}(k) &= \eta I_{g} I_{\text{final}} T_{e}(k) - T_{b}(k), \\
\omega_{w}(k) &=\frac{\omega_{e}(k)}{I_{g} I_{\text{final}}},
\end{aligned}
\end{equation}
where $T_{\text{final}}$, $T_{e}$, $T_{b}$ are the torque of final driveshaft, engine, and brake, respectively, $\omega_{e}$ is the rotational velocity of engine, $\eta$ is the transmission efficiency, $I_{g}$, and $I_{\text{final}}$ are the gear ratios of gearbox and final driveshaft, respectively.

\subsubsection{Fuel Cut-Off}

\begin{figure}
\begin{center}
\includegraphics[width=3.2 in]{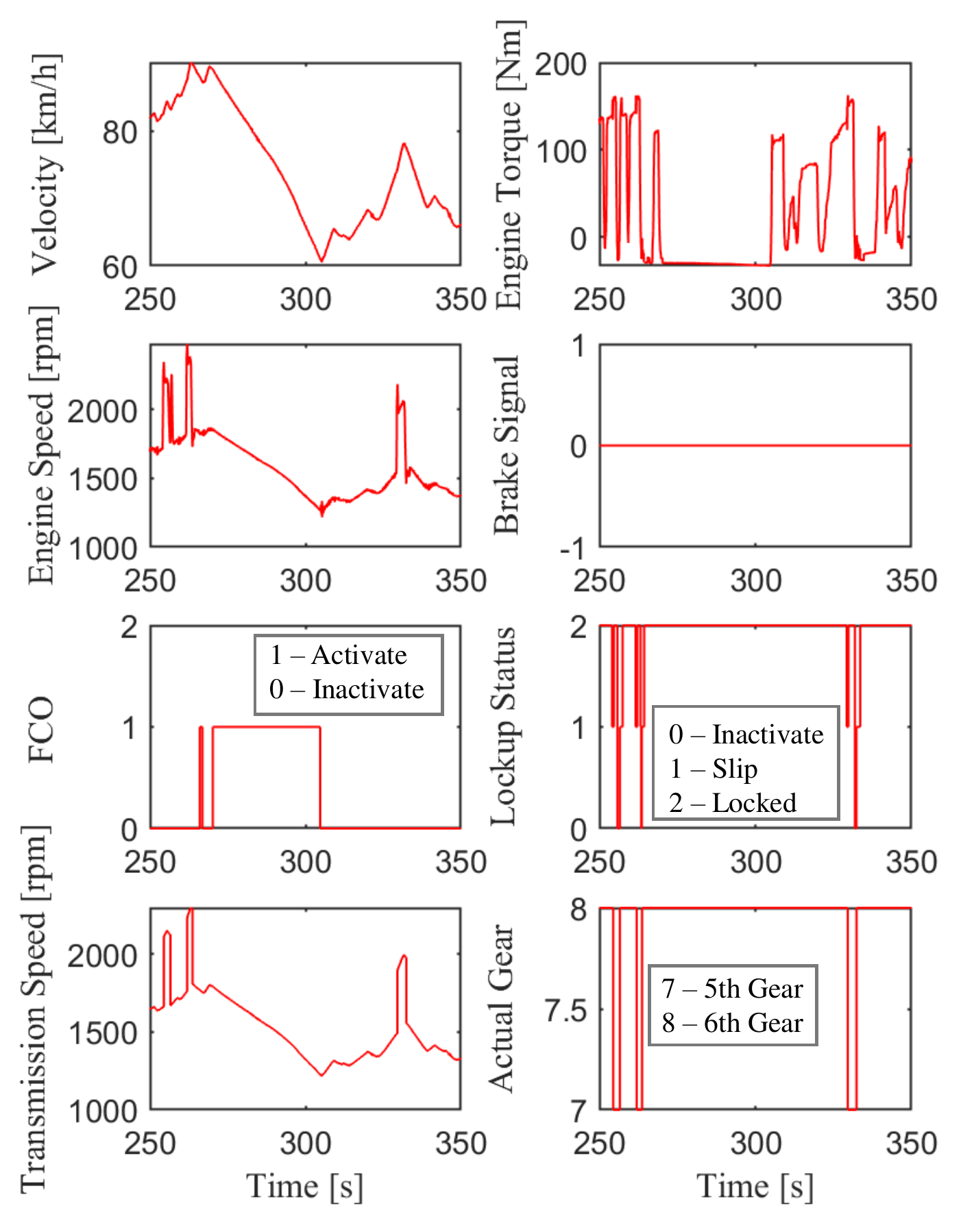}
\caption{Field test data in Wuhan.}
\label{Fig2}
\end{center}
\end{figure}

The fuel cut-off (FCO) strategy, as an efficient way to reduce fuel consumption, has been applied in production engines. As an example, Fig.~\ref{Fig2} shows vehicle driving data collected on a test platform, FCO is enabled when the request engine torque is zero. For FCO, the clutch is engaged and holds the transmission in the highest gear while coasting. The negative engine torque in Fig.~\ref{Fig2}, measured based on the experimental data, indicates the drag.

Following the analysis of field test data, the coasting dynamic with FCO mechanism in the distance domain after the discretization is modeled as
\begin{equation}
\begin{aligned}
T_{\text{final},2}(k) = \eta I_{g}I_{\text{final}} \left(T_{e}(k)-(1-z) T_{\text {drag }}\right)-T_{b}(k)
\end{aligned}
\label{Discrete dynamic model of FCO}
\end{equation}
where $z \in\{0,1\}$ is FCO signal, with 0 and 1 representing the activation and inactivation of fuel cut-off respectively. $T_{\text {drag }}$ is a constant representing the engine drag torque when fuel is cut off. 

The fuel consumption model is modified as follow:
\begin{equation}
\label{piece-wise-function}
M_{f}(k)=\left\{\begin{array}{cl}
\frac{\dot{m}_{f}\left(T_{e}(k)\right)}{v(k)} \Delta s, & z(k)=1 \quad (\text{fuel injected}), \\
0, & z(k)=0 \quad (\text{fuel off}).
\end{array}\right.
\end{equation}
(\ref{piece-wise-function}) reflects that there is no idle fuel consumption when fuel is cut off.

\subsubsection{Engine Start/Stop}

For a conventional vehicle, another fuel efficient strategy during coasting is to turn off the engine and disengage the clutch. Then, the engine will be restarted leveraging the slipping of the clutch powered by vehicle inertia. The clutch start method is superior to belt-driven starter systems or electric motors mounted on the engine's crankshaft in the high speed cruising condition~\cite{mueller2011next}. For the sake of simplicity, “Engine Start/Stop” refers to “Engine Start/Stop coordinated with the engagement or disengagement of the lock-up clutch”.The powertrain model is formulated as
\begin{equation}
\label{dynamic equation of engine start/stop}
\begin{aligned}
T_{\text{final},3}(k) = \eta I_{g}I_{\text{final}} T_{e}(k)-T_{b}(k),
\end{aligned}
\end{equation}
the engine drag torque is eliminated thanks to the disengaged clutch.

Once the engine is turned off and the clutch is disengaged, the rotational speed of the engine will reduce to zero and the rotational speed of the transmission will be proportional to the vehicle speed due to the mechanical connection. To achieve a smooth re-engagement process, the transmission is assumed to stay in the highest gear to minimize the speed difference of the clutch. High-speed cruise conditions normally call for the highest gear for efficient engine operation area. Therefore, we assume that the gear is constant during the process. This assumption is validated by the experiment date in Fig.~\ref{Fig2}.

This restart process can be regarded as using the vehicle inertial energy to “charge” the engine. The extra cost used to restart the engine is modeled based on the energy balance
\begin{equation}
\begin{gathered}
\Delta E_{\text {engine }}=\frac{1}{2} I_{\text {engine }}\left(\omega_{e}^{2}(k)-0\right) \\
\omega_{e}(k)=\frac{ I_{g} I_{\text{final}} v(k)}{2 \pi r_{w}},
\end{gathered}
\end{equation}
where $\Delta E_{\text {engine }}$ is the energy required for cranking the engine from rest to the synchronization speed of transmission. $I_{\text {engine }}$ is the engine inertia. $\omega_{e}(k)$ is the engine speed after synchronization, which is proportional to the vehicle speed due to the engaged lock-up clutch. In this study, it is assumed that the engine start/stop and clutch engage/disengage process is finished within one sampling interval. When the engine is turned on at the $k$th step. According to the energy balance, the partial inertia energy of the vehicle will be transformed into engine inertial energy. This transformation is formulated as 
\begin{equation}
\begin{aligned}
\frac{1}{2} m_{\text{eff}}^{2} v(k+1)= & \frac{1}{2} m_{\text{eff}}^{2} v(k) + \left[F_{t}-F_{r}\right] \Delta s+\delta E_{v}(k),
\end{aligned}
\end{equation}
\begin{equation}
\Delta E_{\text {engine }}(k)=\delta E_{v}(k)=\frac{1}{2} m_{\text{eff}}\left(\bar{v}^{2}(k+1)-v^{2}(k)\right),
\end{equation}
where $\delta E_{v}$ is the partial inertia energy of the vehicle and $\bar{v}(k+1)$ is the partial velocity variation caused by the engine start process within this interval. Thus, the velocity reduction caused by the engine start is formulated as 
\begin{equation}
\Delta v (k)=\left\{\begin{array}{cc}
\bar{v}(k+1)-v(k) & \text { if } d(k)=1 \text { and } d(k-1)=0 \\
0 & \text { else }.
\end{array}\right.
\end{equation}
where $d \in\{0,1\}$ is the engine start/stop signal, with 0 and 1 representing that the engine is turned off and turned on respectively.

The coasting dynamic with engine start mechanism in the distance domain after the discretization can be expressed as
\begin{equation}
\begin{split}
\label{dynamic equation of engine start/stop vehicle level}
v(k+1) & =g\left(v(k), T_{e}(k), T_{b}(k)\right)\\
& =v(k)+\frac{1}{m_{\text{eff}} v(k)}\left[F_{t}-F_{r}\right] \Delta s - \Delta v (k).
\end{split}
\end{equation}

\section{Offline evaluation of eco-coasting methods}
\label{DP OFFLINE EVALUATION}
\label{Section3}
In this section, optimization problems regarding the three coasting methods, including the baseline one, are summarized. Then, the performance of these coasting strategies is evaluated using DP under different slope profiles from the digital map. As we are mostly concerned with fuel economy and time spent in traveling, fuel consumption and travel time are used in the cost function and the weighted sum method is used to evaluate the trade off between these competing objectives.

\subsection{Fuel-efficient Optimal Control of Baseline Method}
Given a priori slope profile with a length of $N$, an optimization problem is formulated to find an optimal control law, that minimizes the cost function including fuel consumption and travel time over the entire driving cycle,
\begin{equation}
\min _{T_{e}(.), T_{b}{(.)}} \sum_{i=0}^{N}(\beta ({M}_{f}(k))+(1-\beta) \frac{1}{v(k)}) \Delta s
\end{equation}
subject to the system dynamic constraints (\ref{Discrete dynamic vehicle-level model}), (\ref{dynamic equation of baseline model}) and 
\begin{equation}
\label{Constraints of baseline}
\begin{aligned}
& 0 \leq T_{e}( k) \leq T_{e, \max }, \quad k=0: N-1\\
& 0 \leq T_{b}( k) \leq T_{b, \max }, \quad k=0: N-1\\
&v_{\min } \leq v(k) \leq v_{\max }, \quad k=0: N \\
&v(0)=v_{0}, v(N)=v_{0},
\end{aligned}
\end{equation}
where $T_{e, \max }$ are the engine torque constraints, $T_{b, \max }$ are the brake torque constraints. $v_{\min }$ and $v_{\max }$ are the lower and upper limits of cruise speed, implying that the vehicle maintains the highest gear within this speed range. $\beta$ is the weight ranging from 0 to 1. The initial and terminal speed is fixed as $v_{0}$ to make a fair comparison. \\

\subsection{Fuel-efficient Optimal Control of FCO Method}
To avoid a high switching frequency of the FCO signal, which is associated with driving comfort, an additional penalty is added to the cost function
\begin{equation}
\label{cost function of DP}
\begin{aligned}
\min _{T_{e}(\cdot), T_{b}(\cdot), z(\cdot)} \sum_{i=0}^{N}(\beta M_{f}(k) & + (1-\beta) \frac{1}{v(k)}\\
& +\alpha \cdot(z(k+1)-z(k))^{2}) \Delta s
\end{aligned}
\end{equation}
where the added term $ \cdot(z(k+1)-z(k))^{2}$ reflects the FCO switch frequency during the whole trip. With a weighted coefficient $\alpha$, this term limits the switching numbers of the FCO signal. The cost of (\ref{cost function of DP}) subject to the system dynamics (\ref{Discrete dynamic vehicle-level model}), (\ref{Discrete dynamic model of FCO}) and 
\begin{equation}
\begin{aligned}
\label{Constraints of FCO}
&0 \leq T_{e}(k) \leq T_{e, \max } \cdot z(k), \quad k=0: N-1\\
&0 \leq T_{b}(k) \leq T_{b, \max } \cdot(1-z(k)), \quad k=0: N-1 \\
&v_{\min } \leq v(k) \leq v_{\max }, \quad k=0: N \\
&v(0)=v_{0}, v(N)=v_{0} \\
&z(k) \in\{0,1\}, \quad k=0: N-1.
\end{aligned}
\end{equation}

The constraints above imply that $T_{e}(k)=0$ when FCO is activated, and $T_{b}(k)=0$ when FCO is inactivated. This condition automatically assures that $T_{e}$ and $T_{b}$ will not be non-zero simultaneously.

\subsection{Fuel-efficient Optimal Control of Engine Start/Stop Method}
Since the cost to restart the engine is formulated in the dynamic equation (\ref{dynamic equation of engine start/stop}) and (\ref{dynamic equation of engine start/stop vehicle level}), the penalty term $ (d(k+1)-d(k))^{2}$ intended to limit frequent switch is redundant and the cost function for engine start/stop control is comprised of fuel consumption and travel time as same as the baseline, 
\begin{equation}
\begin{aligned}
\min _{T_{e}(\cdot), T_{b}(\cdot), d(\cdot)} \sum_{i=0}^{N}(\beta M_{f}(k) & + (1-\beta) \frac{1}{v(k)}) \Delta s
\end{aligned}
\end{equation}
subject to the system dynamics (\ref{dynamic equation of engine start/stop}), (\ref{dynamic equation of engine start/stop vehicle level}). The same state variable and control variable constraints as (\ref{Constraints of FCO}) for the fuel cut-off method are imposed to the optimization problem formulation of the engine start/stop method, except that the fuel cut-off signal $z(k)$ in (\ref{Constraints of FCO}) is replaced by the engine start/stop signal $d(k)$.

\subsection{Dynamic Programming}
To compare the fuel saving benefit among different eco-coasting strategies, Dynamic Programming~\cite{hellstrom2010design}, as a global optimization approach, is used in this research to derive the timing and duration of the coasting signal along with the engine torque and brake torque.

The optimal control law is solved using backward iteration by minimizing the cost-to-go function.
In this study, the discretization distance is 5 $m$, the other parameters corresponding to constraints are listed in Table~\ref{Parameters of optimization problem}. Note that the maximum of the brake torque is set to 500 $Nm$, rather than the actual upper limits as specified by the brake system characteristics, to avoid unexpected sudden brake during this cruising condition. 

\begin{table}[htbp]
\centering
\caption{Parameters of optimization problem}
\label{Parameters of optimization problem}
\begin{tabular}{c c c}
\toprule
Parameter  & Description  & Value\\\hline
$v_{\min }$         &  Minimum of velocity          & 50 $km/h$   \\
$v_{\max }$        &  Maximum of velocity           & 90 $km/h$\\
$T_{e, \max }$    &  Maximum of engine torque       & 120 $Nm$ \\
$T_{b, \max }$   &  Maximum of brake torque         & 500 $Nm$\\
$v_{0}$        &  Initial and terminal velocity     & 75 $m/s$ \\
\bottomrule
\end{tabular}
\end{table}

\subsection{Evaluation and Discussion}
\label{Offline Evaluation Results and Discussion}

\begin{figure}
\begin{center}
\includegraphics[width=3.2 in]{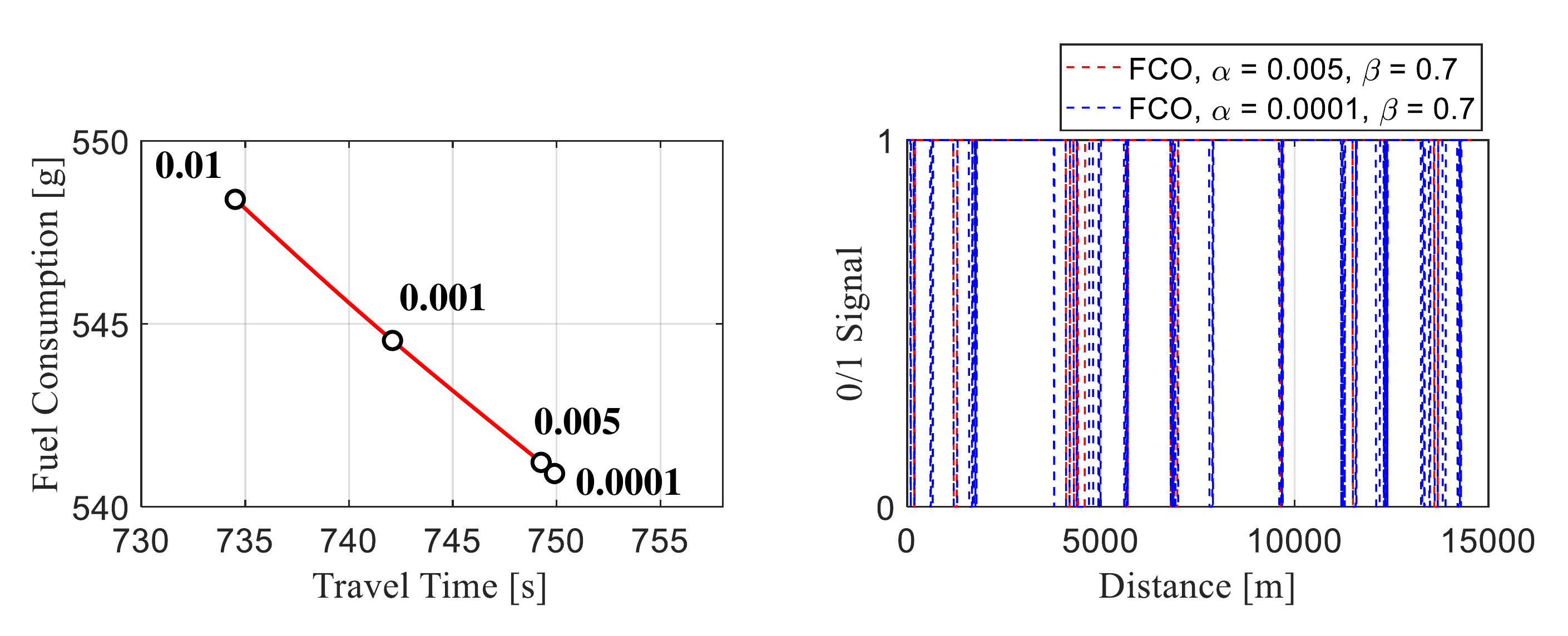}
\caption{The performance trade off with FCO coasting method by varying $\alpha$ (left). FCO signal with different $\alpha$ under Wuhan slope profile (right).}
\label{Fig3}
\end{center}
\end{figure}

\begin{figure*}
\begin{center}
\includegraphics[width=6 in]{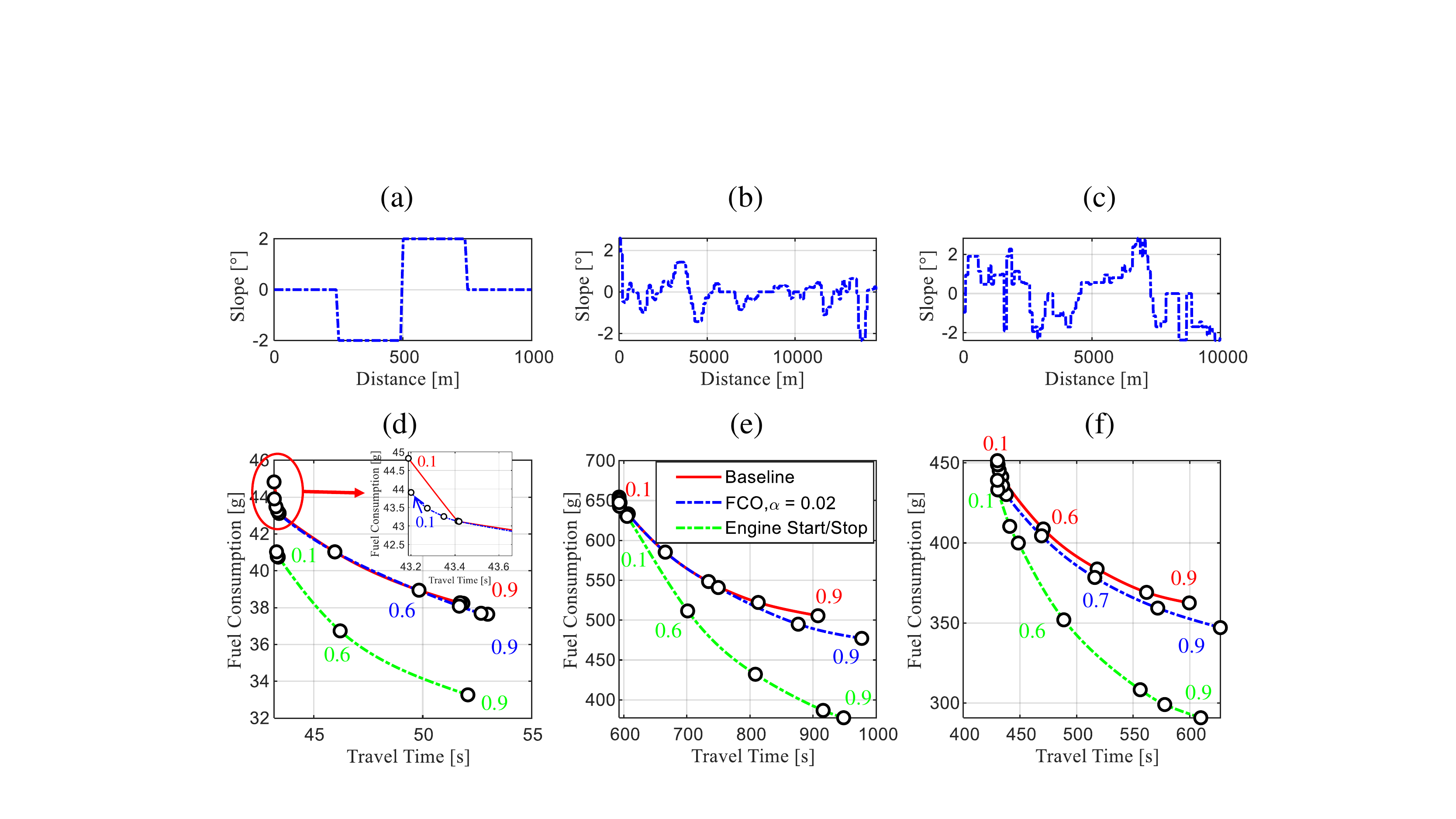}
\caption{The Pareto Front with baseline, FCO, and engine start/stop coasting methods under different slope profile: (a) and (d) The standard uphill-downhill slope profile and the corresponding Pareto Front, (b) and (e) The Chongqing slope profile and the corresponding Pareto Front, (c) and (f) The Wuhan slope profile and the corresponding Pareto Front. }
\label{Fig4}
\end{center}
\end{figure*}

\begin{figure*}
\begin{center}
\includegraphics[width=5 in]{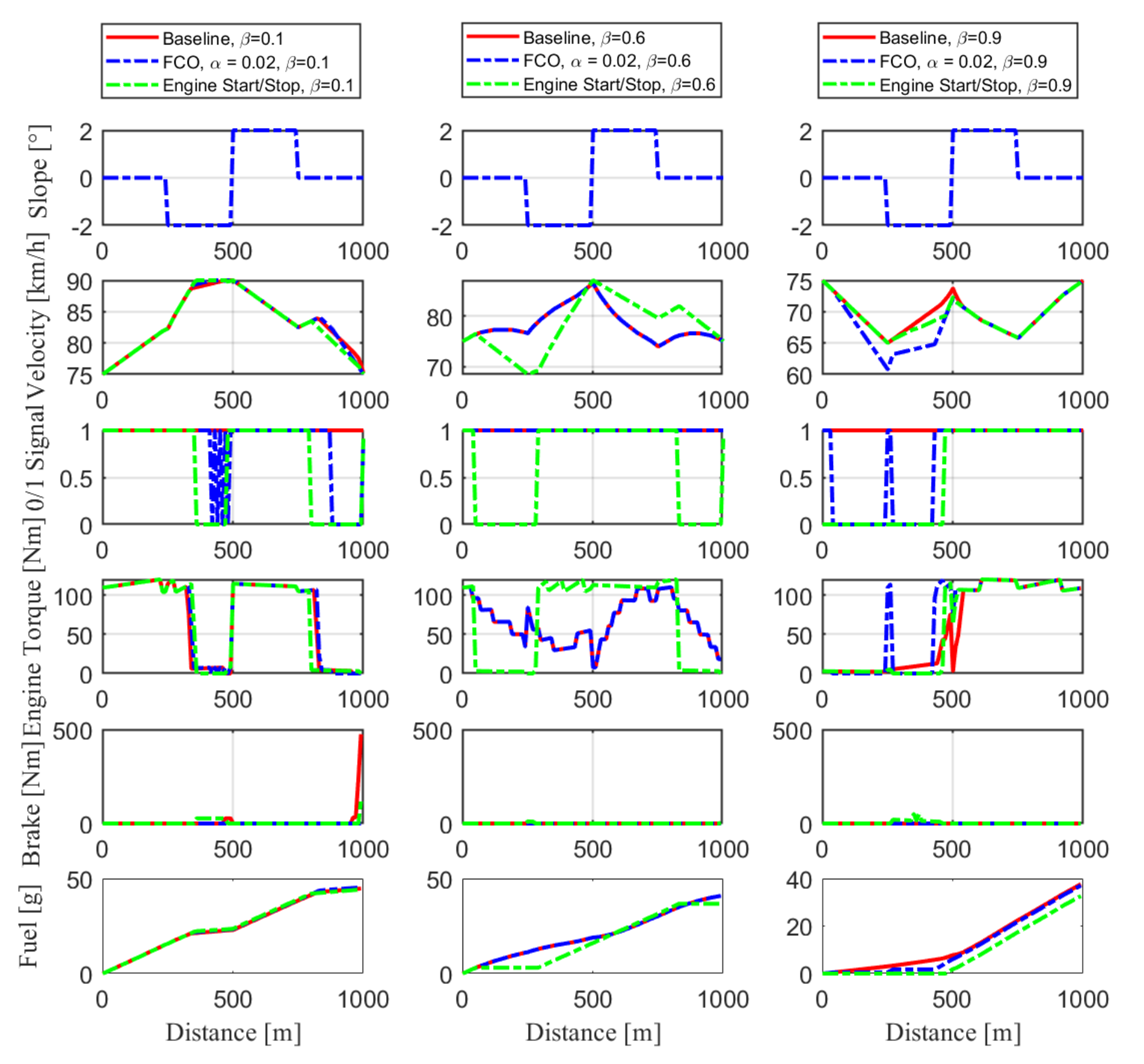}
\caption{DP simulation results under standard uphill-downhill slope profile with $\beta=0.1$, $\beta=0.6$, and $\beta=0.9$.}
\label{Fig5}
\end{center}
\end{figure*}

\begin{figure}
\begin{center}
\includegraphics[width=2.5 in]{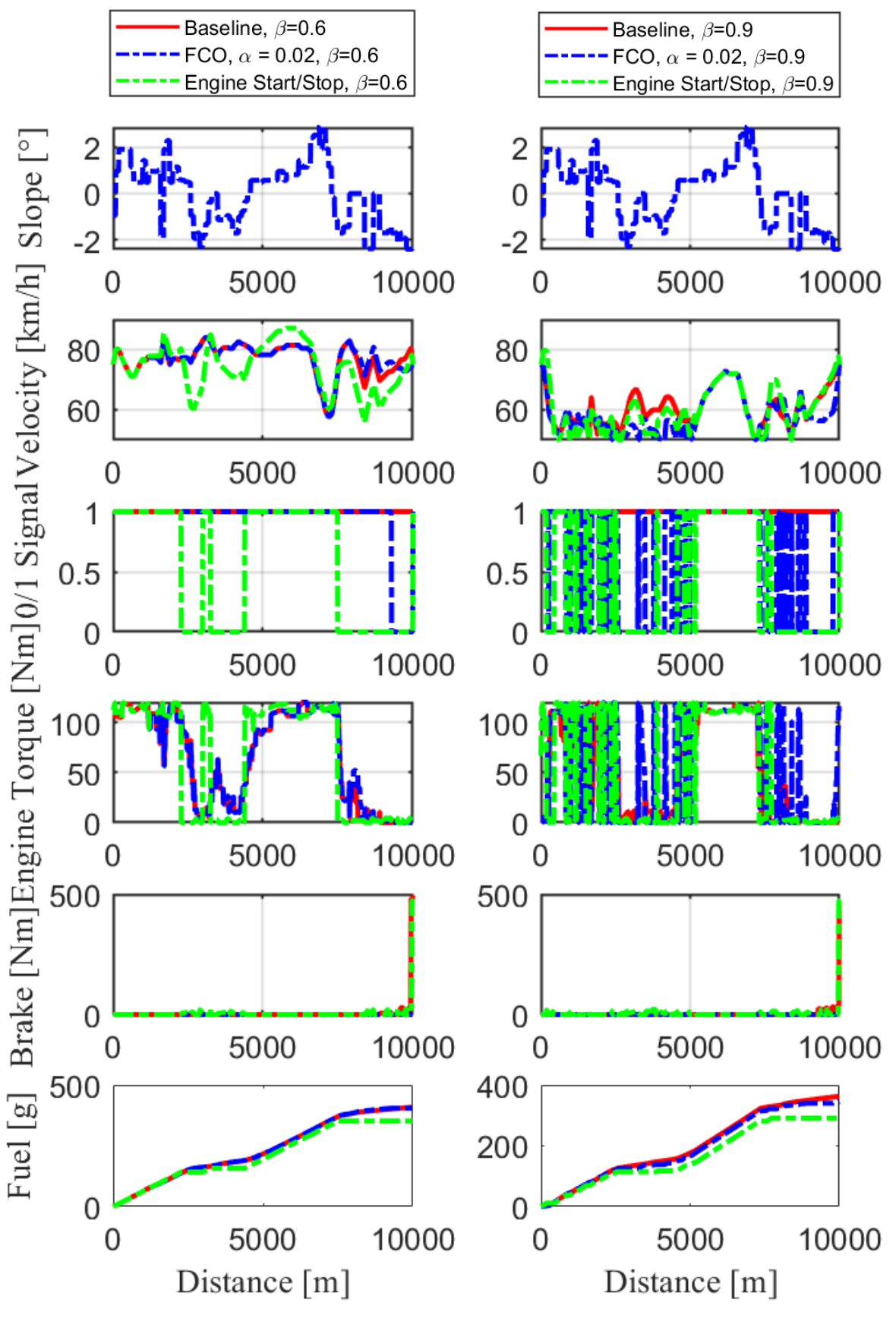}
\caption{DP simulation results under Chongqing slope profile with $\beta=0.6$, and $\beta=0.9$.}
\label{Fig6}
\end{center}
\end{figure}

In this section, three slope profiles are used to assess the performance metrics of different coasting strategies with DP algorithm. The first slope profile is a typical uphill-downhill segment for 1000 m as shown in Fig.~\ref{Fig4}. The other two profiles come from measurements on Chongqing and Wuhan highway separately. There is no single solution that would simultaneously optimize all criteria which are naturally in conflict. Varying the weighting factor $\beta$ allows us to put a different relative emphasis on each attribute to investigate performance trade-offs (i.e., the Pareto front). The Pareto fronts provide insights into the trade-off between travel time and fuel consumption for different coasting methods. In such case, none of solutions from the Pareto front can be improved, in the sense that improving one attribute will lead to deterioration of the other~\cite{sadek2016fpga}.

Based on the simulation results in Fig.~\ref{Fig3}, for the FCO method, more FCO mechanism is activated when the $\alpha$ is assigned as a larger value consuming less fuel but extending the time. 

In terms of travel time and fuel consumption as shown in Fig.~\ref{Fig4}, the baseline strategy and FCO strategy have the same performance. The Engine start/stop strategy is outperformed than the other two method due to the fact that the performance of baseline strategy and FCO strategy is dominated by many solutions from the engine start/stop strategy. The reason is that, as shown in Fig.~\ref{Fig5} with $\beta=0.1$, the vehicle will turn off the engine to coast in advance considering the terminal speed constraints. The disengagement of the lock-up clutch during the engine off period can eliminate the engine drag torque. Therefore, the decoupled engine and transmission yield a moderate deceleration and a longer coasting distance consequently. The longer coasting distance, compared with the FCO approach, can decrease fuel consumption without compromising travel time dramatically. This pattern is also observed when the weight turns to 0.6 and 0.9, as shown in Fig.~\ref{Fig5}. 

This optimization result leverages engine-off condition under downhill, leading to better performance. Based on the simulation results under the Chongqing slope profile with $\beta=0.6$ (Fig.~\ref{Fig6}), the engine-off is activated in almost every downhill. In comparison, to reduce idle fuel consumption, the fuel cut-off condition is only activated at the end of the driving cycle. Furthermore, the terminal speed constraint is satisfied as well thanks to the engine drag torque under the downhill condition. 

Under Chongqing and Wuhan slope profiles, the FCO and engine start/stop coasting methods can achieve better performance with increased weight $\beta$, indicting that fuel consumption will be emphasized more in the cost function and this will induce more fuel cut-off and engine-off conditions, as illustrated in Fig.~\ref{Fig6}.

The cost for restarting the engine is included in the dynamic equation (\ref{dynamic equation of engine start/stop vehicle level}) and will reduce the switching frequency of the engine start/stop signal. As shown in Fig.~\ref{Fig6}, this formulation will not only provide a fair evaluation of the overall performance (i.e., fuel consumption and travel time), but also eliminate the driving discomfort associated with frequent start/stop.

\section{Online Implementation with MPC}

Based on the DP simulation results of Section~\ref{DP OFFLINE EVALUATION}, the engine start/stop coasting strategy has an obvious advantage compared with the other two methods. Compared with the baseline, FCO provides moderate improvement with significant emphasis put on fuel consumption. However, this offline calculated solution is not feasible for online implementation as it requires future information of the driver's inputs. To develop an online solution, MPC is developed for these two proposed coasting methods and performance is evaluated. To better assess the performance of these two coasting approaches with different control methods (e.g., PI controller, MPC, DP), a reference speed is extracted from the in-field experiment in which the vehicle is controlled by a human driver~\cite{chu2018predictive}. In this section, the previous travel time term in the cost function is replaced by the tracking error. The cost function of DP for FCO and engine start/stop coasting strategies are reformulated as
\begin{equation}
\begin{aligned}
\min _{T_{e}(\cdot ), T_{b}(\cdot ), z(\cdot )} \sum_{i=0}^{N}(& \beta \cdot\left(M_{f}(k)\right)+\\
& (1-\beta) \cdot\left(v( k)-v_{\text{ref}}\right)^{2}) \Delta s,
\end{aligned}
\end{equation}
and 
\begin{equation}
\begin{aligned}
\min _{T_{e}(\cdot), T_{b}(\cdot), z(\cdot)} \sum_{i=0}^{N}(\beta \cdot M_{f}(k) & + (1-\beta) \cdot\left(v( k)-v_{\text{ref}}\right)^{2}\\
& +\alpha \cdot(z(k+1)-z(k))^{2}) \Delta s.
\end{aligned}
\end{equation}

In this section, a rule-based approach is first designed to provide a reference. Then, a speed-tracking MPC formulation is proposed for these two coasting strategies. A novel constraint is applied to satisfy the minimum off steps constraints of the MIMPC and mitigate the switching frequency. After that, the performance is discussed by comparing with the PI controller and DP simulation results. Moreover, the effect of the length of the prediction horizon is demonstrated. Finally, to mitigate the computation burden, a heuristic MPC is proposed by using the engine start/stop sequence obtained from the offline DP results.

\subsection{PI Controller}

\begin{figure}
\begin{center}
\includegraphics[width=3.2 in]{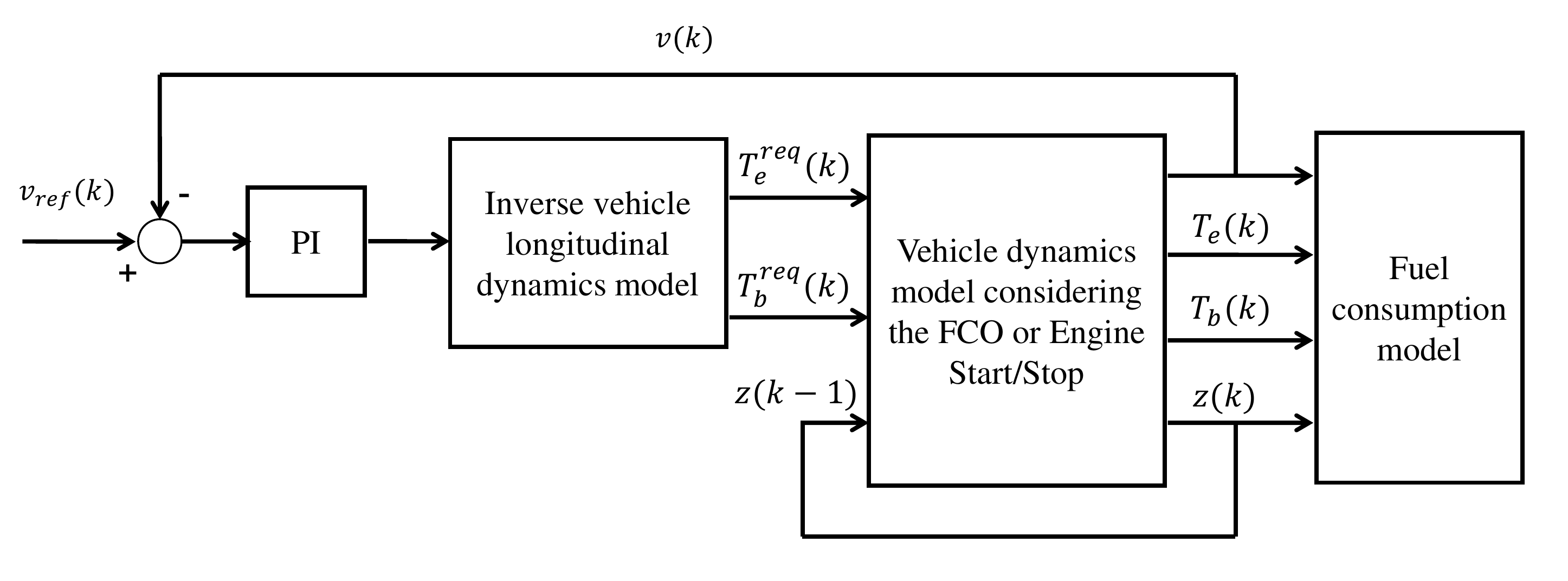}
\caption{The block diagram of PI controller. $T_{e}^{r e q}$ is the required engine torque, $T_{b}^{r e q}$ is the required brake torque, $z(k-1)$ is the binary variable at the last step.}
\label{Fig7}
\end{center}
\end{figure}

FCO is available in our in-field experiment platform in this study~\cite{chu2018predictive}. It is activated when no torque is requested and the speed is higher than a threshold. Inspired by this trigger condition, a rule-based FCO or Engine-turn-off mechanism is proposed. The FCO or Engine turn-off signal is activated if the current speed is higher than $75km/h$ and the required engine torque is decreased to zero simultaneously. A PI controller is applied to track the reference speed, as shown in Fig.~\ref{Fig7}.

The FCO or engine turn-off signal will hold for 4 discretization steps if it is activated to avoid frequent switches, similar to the engine on command filter in~\cite{hou2014approximate}. The operational constraints on minimum off discretization steps chosen in this study is tuned offline considering the trade-off between tracking accuracy and fuel consumption.

\subsection{MPC}

MPC, also known as the receding horizon control, is a control technique that embeds optimization within feedback to deal with systems subject to constraints on inputs and states~\cite{li2009path}. MPC determines the control action by solving a finite horizon open-loop optimal control problem online at each discrete instant. However, in this study, there exists discrete dynamics in the engine start/stop or FCO coasting methods, resulting in a mixed-integer optimal control problem (MIOCP). 

Frequent switch of the engine start/stop or fuel cut-off signal will affect driving comfort and cause carbon deposits in the engine. To avoid that, the operational constraints on the minimum amount of steps for which an engine or fuel injection must be kept off can be expressed by the following mixed integer linear inequalities:
\begin{equation}
\delta(\tau) \leq 1-(\delta(j-1)-\delta(j))
\end{equation}
with $\tau=j+1, \ldots, \min \left(j+d_{\min }-1,N\right)$, $d_{\min}=4$ represents the operational constraints on minimum off steps, $j=2, \ldots, N+d_{\min }$. Let the extended state be represented by the vector $\delta:=\left[z_{\text {history }}, z_{\text {current }}\right]^T$, where  $z_{\text {history }}$ is the vector of the previous binary variables with dimension $d_{\min } \times 1$ as
\begin{equation}
\begin{aligned}
z_{\text {history }}:=\left[z_{(k-d_{\min }-1)}, \ldots, z_{k-1}\right]^T.
\end{aligned}
\end{equation}
$z_{\text {current}}$ is the row vector of the current binary variables with dimension $N_{h} \times 1$ as
\begin{equation}
\begin{aligned}
z_{\text {current}}:=\left[z_{1|k}, \ldots, z_{N_{h|k}}\right]^T.
\end{aligned}
\end{equation}

The implementation of the mixed integer linear inequalities in the MPC formulation is shown in Algorithm~\ref{alg:ALG1}. More details about the constraints can be found in~\cite{parisio2016stochastic} and~\cite{carrion2006computationally}.

\begin{algorithm}
        \caption{MIMPC with minimum off constraints} 
        \label{alg:ALG1}
        \begin{algorithmic}[1]
 
        \State Initialize $v_{0|k}$, $C$ and $z_{\text {history }}$; 
        \For {$j=1,\ldots, N_{h}$} \do\\
        \For {$i=1,\ldots, (d_{\dim}+N_{h})$}
        \State $C:=[C, \tau(k:\min(N_{h},i+d_{\dim}-1)) \leq \tau(i)-\tau(i-1)]^T$
        \EndFor
        \State Impose constraints vector $C$ to original MIOCP
        \State Calculate $T_{e} \in \mathcal{R}^{N_{h}}$, $T_{b} \in \mathcal{R}^{N_{h}}$, $z \in \mathcal{R}^{N_{h}}$
        \State Update $z_{\text{history}} := \left[z_{j-d_{\min}}, \ldots, z_{k-1}, z_{1|k}\right]^T$
        \EndFor
    \end{algorithmic}
\end{algorithm}

\subsubsection{FCO}
Let $x(i | k)$ denotes the value of variable $x$ at distance $k+i$ while the prediction is made at the current location $k$. The MIMPC for fuel cut-off is defined over a finite-distance horizon ($N_{h}$) as:
\begin{equation}
\begin{aligned}
\min _{T_{e}(\cdot | k), T_{b}(\cdot | k), z(\cdot | k)} \sum_{i=0}^{N_{h}}(& \beta \cdot\left(M_{f}(i | k)\right)+\\
& (1-\beta) \cdot\left(v(i | k)-v_{\text{ref}} \right)^{2}) \Delta s
\end{aligned}
\end{equation}
subject to:\\
\begin{equation}
\begin{aligned}
\label{constraints of FCO}
&v(i+1|k) =v(i|k) +\frac{1}{m_{\text{eff}} v(i|k)}\left[F_{t}-F_{r}\right] \Delta s, i=0: N_{h}-1\\
& 50 km/h \leq v(i | k) \leq 90 km/h,  i=0: N_{h} \\
& 0 Nm \leq T_{e}(i | k) \leq 120 Nm \cdot z(i | k),  i=0: N_{h}-1 \\
& 0 Nm \leq T_{b}(i | k) \leq 500 Nm \cdot(1-z(i | k)),  i=0: N_{h}-1 \\
&z(i | k) \in\{0,1\}, i=0: N_{h}-1 \\
&\delta(\tau) \leq 1-(\delta(j-1)-\delta(j)), j=2, \ldots, N_{h}+d_{\min } \\
&v(0)=75 km/h.
\end{aligned}
\end{equation}

The optimal coasting action is achieved by solving the above minimization problem repeatedly in a receding-horizon manner. 

\subsubsection{Engine Start/Stop}
The MIMPC for engine start/stop method is defined over a finite-distance horizon ($N_{h}$) as:
\begin{equation}
\label{cost of engine start/stop}
\begin{aligned}
\min _{T_{e}(\cdot | k), T_{b}(\cdot | k), z(\cdot | k)} \sum_{i=0}^{N_{h}}(& \beta \cdot\left(M_{f}(i | k)\right)+\\
& (1-\beta) \cdot\left(v(i | k)-v_{\text{ref}}\right)^{2}) \Delta s
\end{aligned}
\end{equation}
subject to the system dynamics:\\
\begin{equation}
\label{constraints of engine start/stop}
\begin{aligned}
&v(i+1|k) =v(i|k)\\
&  +\frac{1}{m_{\text{eff}} v(i|k)}\left[F_{t}-F_{r}\right] \Delta s -\Delta v(i|k), i=0: N_{h}-1.
\end{aligned}
\end{equation}
Moreover, the same state variable and control variable constraints as (\ref{constraints of FCO}) for the FCO method with the MPC framework are imposed to the optimization problem formulation of the engine start/stop method, except that the fuel cut-off signal $z(k)$ in (\ref{constraints of FCO}) is replaced by the engine start/stop signal $d(k)$.

We implement the proposed MIMPC above in YALMIP \cite{lofberg2004yalmip} and solve the corresponding non-convex non-linear optimal programming using the BARON commercial solver \cite{tawarmalani2005polyhedral}. The maximum number of allowable iteration is set as 200 since the difference of lower bound and upper bound can converge to zero in most steps for this Branch $\&$ Bound based solver. If the solver can not find a feasible solution within the max iteration time, the last feasible solution will be applied to the plant to guarantee the recursive feasibility.

\subsection{Performance Evaluation and Discussion}

\begin{figure*}
\begin{center}
\includegraphics[width=6 in]{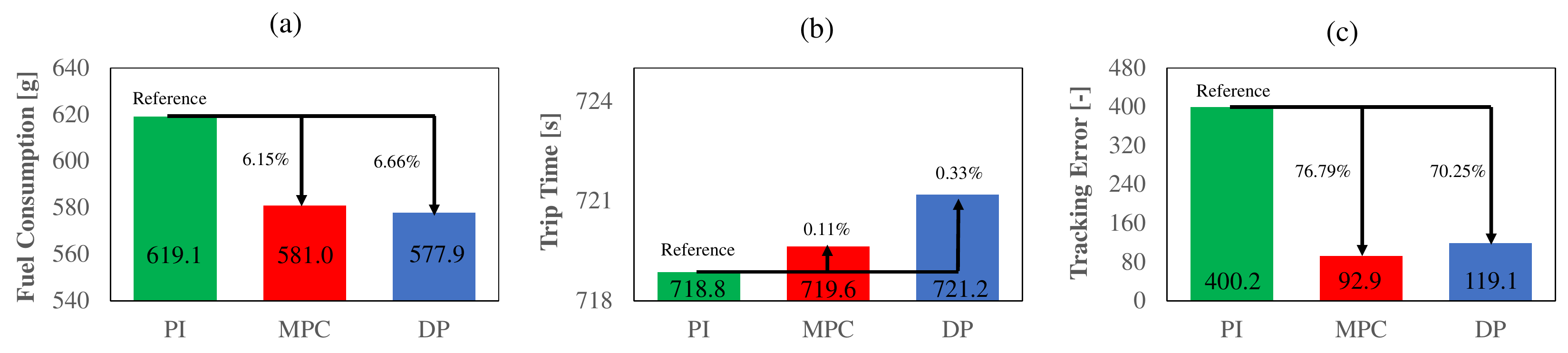}
\caption{Performance comparison with FCO method: DP ($\alpha=0.004,\beta=0.7$) vs MPC ($\beta=0.16$)vs PI controller ($P=0.5, I = 0.0001$).}
\label{Fig8}
\end{center}
\end{figure*}

\begin{figure}
\begin{center}
\includegraphics[width=3 in]{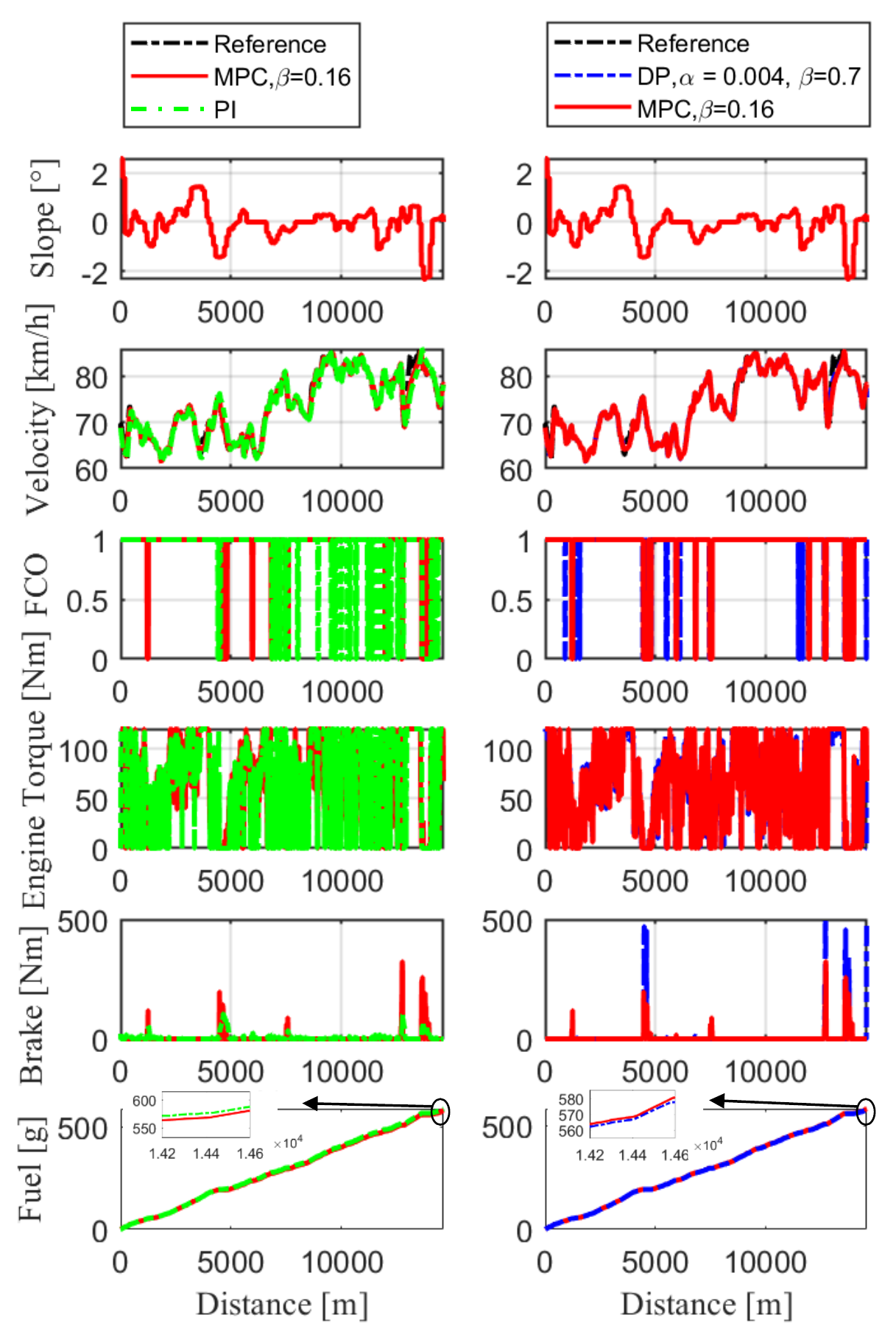}
\caption{Simulation results comparison with FCO method under the Wuhan slope profile.}
\label{Fig9}
\end{center}
\end{figure}

For the FCO coasting method under the Wuhan slope profile, compared with the PI controller, DP achieves a 6.66\% decrease in fuel consumption while the travel time is decreased by $0.33\%$ (Fig.~\ref{Fig8}). The tracking error of the PI controller is higher than DP considering the operational constraints on the minimum off steps. There are two reasons for the fuel consumption reduction: first, the timing and duration of the fuel cut-off results of the rule-based PI controller show a significant difference compared with the global optimal solution from the DP results. Second, more brake torque is involved in the PI controller as shown in Fig.~\ref{Fig9}.

The proposed MPC achieves a substantial performance improvement with respect to the fuel consumption with the FCO coasting method. There is a 6.15\% reduction in fuel consumption compared with the PI controller without compromising the travel time, almost equal to the DP result. As shown in Fig.~\ref{Fig9}, the MPC repeats almost every fuel cut-off action of the global optimal solution but will also take extra fuel cut-off action due to its finite horizon.

\begin{figure*}
\begin{center}
\includegraphics[width=6 in]{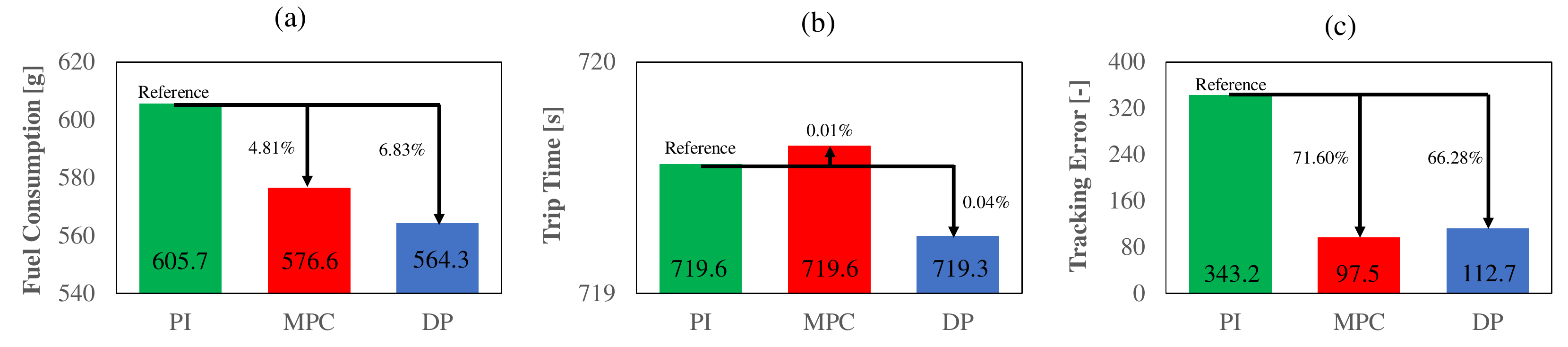}
\caption{Performance comparison with engine start/stop approach: DP ($\beta=0.7$) vs MPC ($\beta=0.16$) vs PI controller ($P=2, I=0.001$).}
\label{Fig10}
\end{center}
\end{figure*}

\begin{figure*}
\begin{center}
\includegraphics[width=6 in]{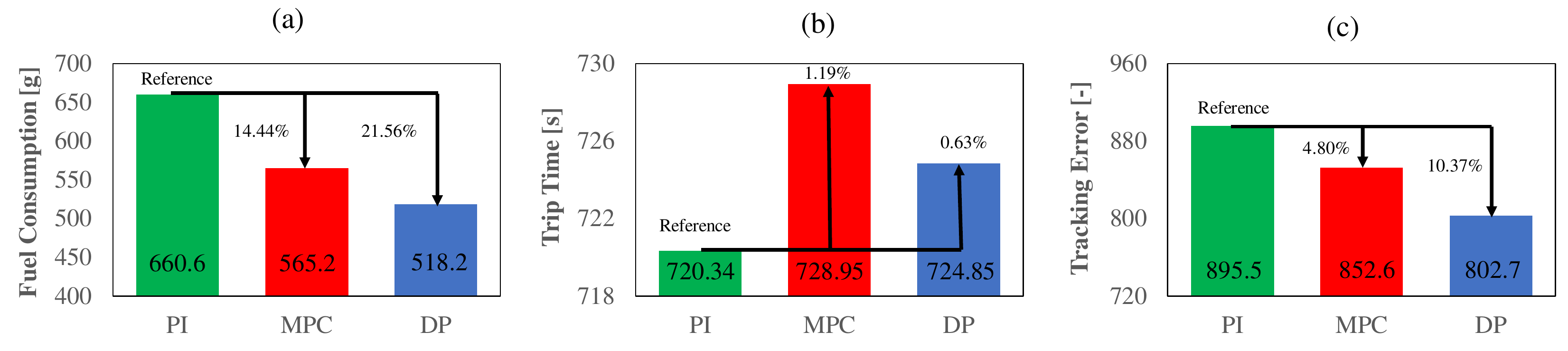}
\caption{Performance comparison with engine start/stop approach: DP ($\beta=0.9985$) vs MPC ($\beta=0.3$) vs PI controller ($P=0.18, I=0.01$).}
\label{Fig11}
\end{center}
\end{figure*}

\begin{figure*}
\begin{center}
\includegraphics[width=5 in]{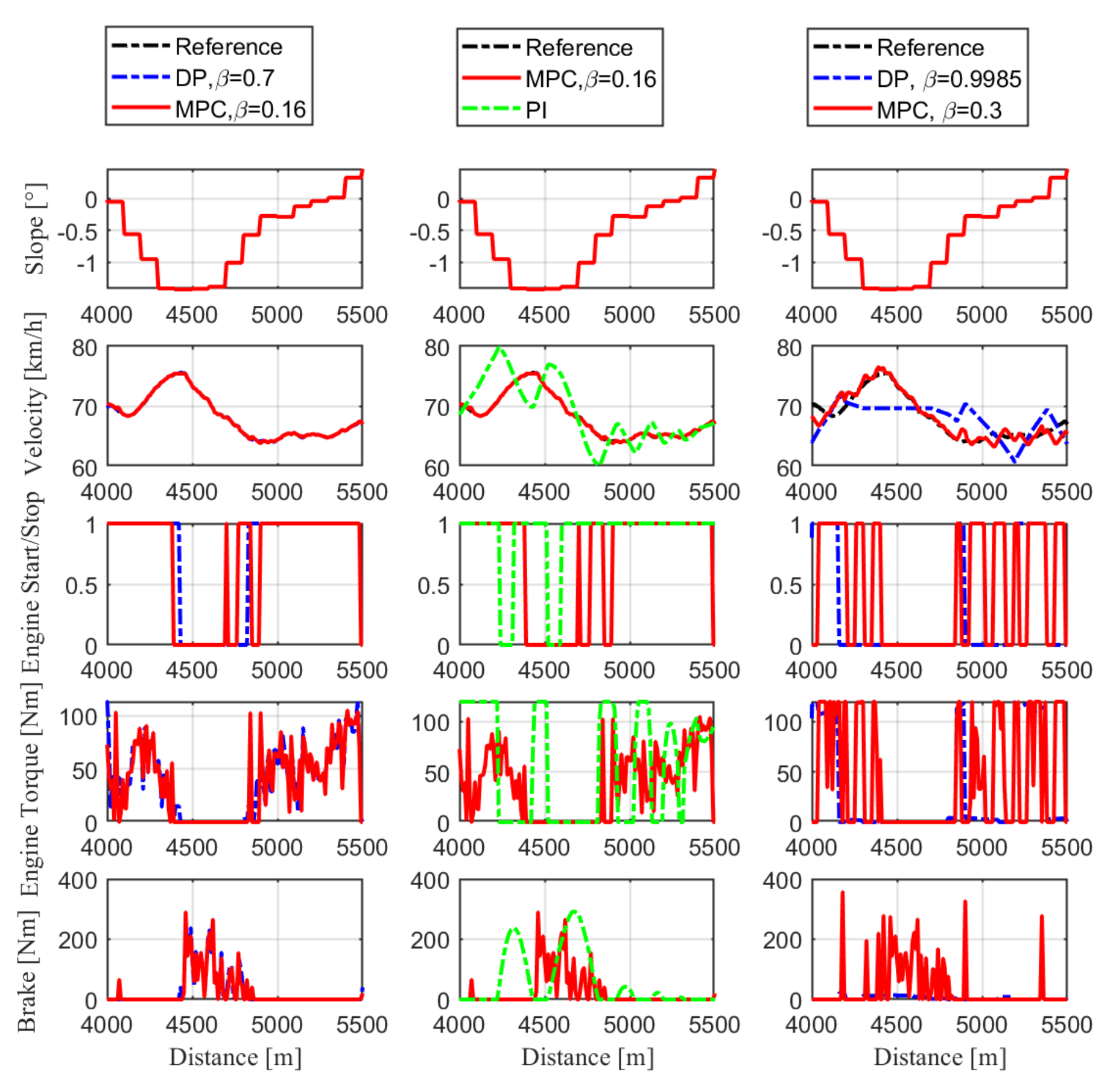}
\caption{Simulation results comparison with engine start/stop method zoomed from 4000 $m$ to 5500 $m$ of the Wuhan slope profile.}
\label{Fig12}
\end{center}
\end{figure*}

For the engine start/stop coasting method, the fuel consumption of MPC is less than the results of the PI controller, close to the DP results. The distinct engine off decisions and duration among these three methods lead to the different performance as shown in Fig.~\ref{Fig10}. As shown in Fig.~\ref{Fig12}, the engine is turned off by the DP at 4500 m as the reference speed starts to decrease at the downhill condition. The brake torque is involved to track the reference speed precisely. MPC also predicts this engine-off action exactly, but this engine start/stop signal switches multiple times when the prediction horizon is chosen as 200 m. Regarding the results of the PI controller at the same spot, as shown in Fig.~\ref{Fig12}, the engine is turned off for a while when its speed is higher than 75 km/h. After that, the engine is turned on referring to its rule-based trigger condition and fuel injection continues to maintain engine idle speed.

Compared with the MPC performance between FCO (Fig.~\ref{Fig8}) and engine start/stop (Fig.~\ref{Fig10}) coasting methods, the FCO has a better performance in terms of the fuel consumption reduction compared with the results of PI controller. However, if the tracking error of the engine start/stop method is relaxed to the same level as the FCO method (Fig.~\ref{Fig11}), there will be a significant improvement with respect to the fuel consumption for MPC. This observation implies that the potential benefit from different coasting mechanisms will be eliminated without proper speed tracking. As shown in Fig.~\ref{Fig12}, DP with $\beta=0.9985$ calculates a longer engine-off period at the cost of a larger tracking error compared with the DP results with $\beta=0.7$ in Fig. 12 at the same location. More engine-off action is calculated to minimize the fuel consumption while sacrificing the tracking accuracy a little bit.

\subsection{Sensitivity Analysis on the Length of the Prediction Horizon}

\begin{figure}
\begin{center}
\includegraphics[width=3.2 in]{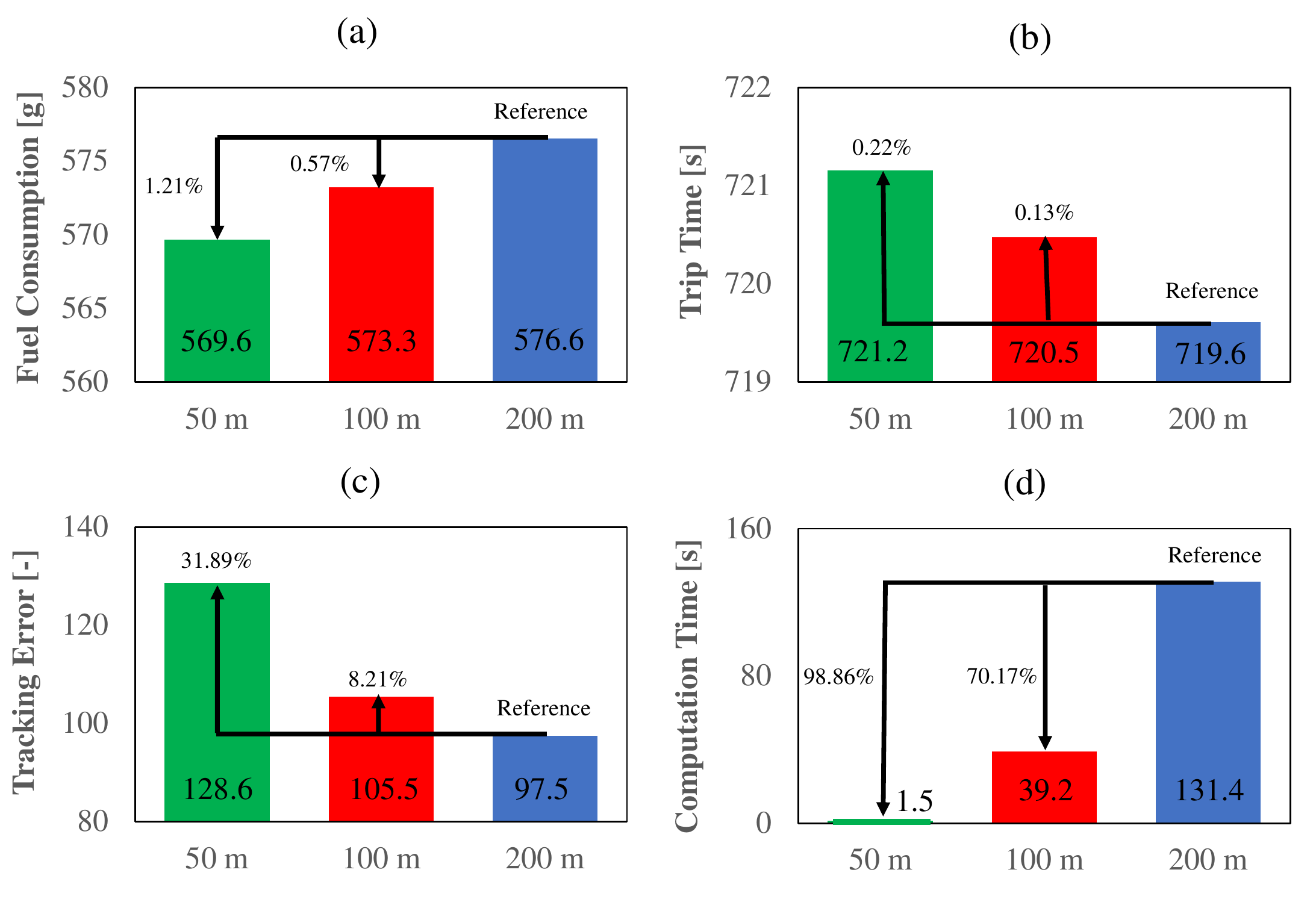}
\caption{Effect of increasing the prediction horizon with engine start/stop method. (a) The fuel consumption; (b) The travel time; (c) The tracking error; (d) The computation time for obtaining our numerical solution at each instant}
\label{Fig13}
\end{center}
\end{figure}

\begin{figure}
\begin{center}
\includegraphics[width=3.2 in]{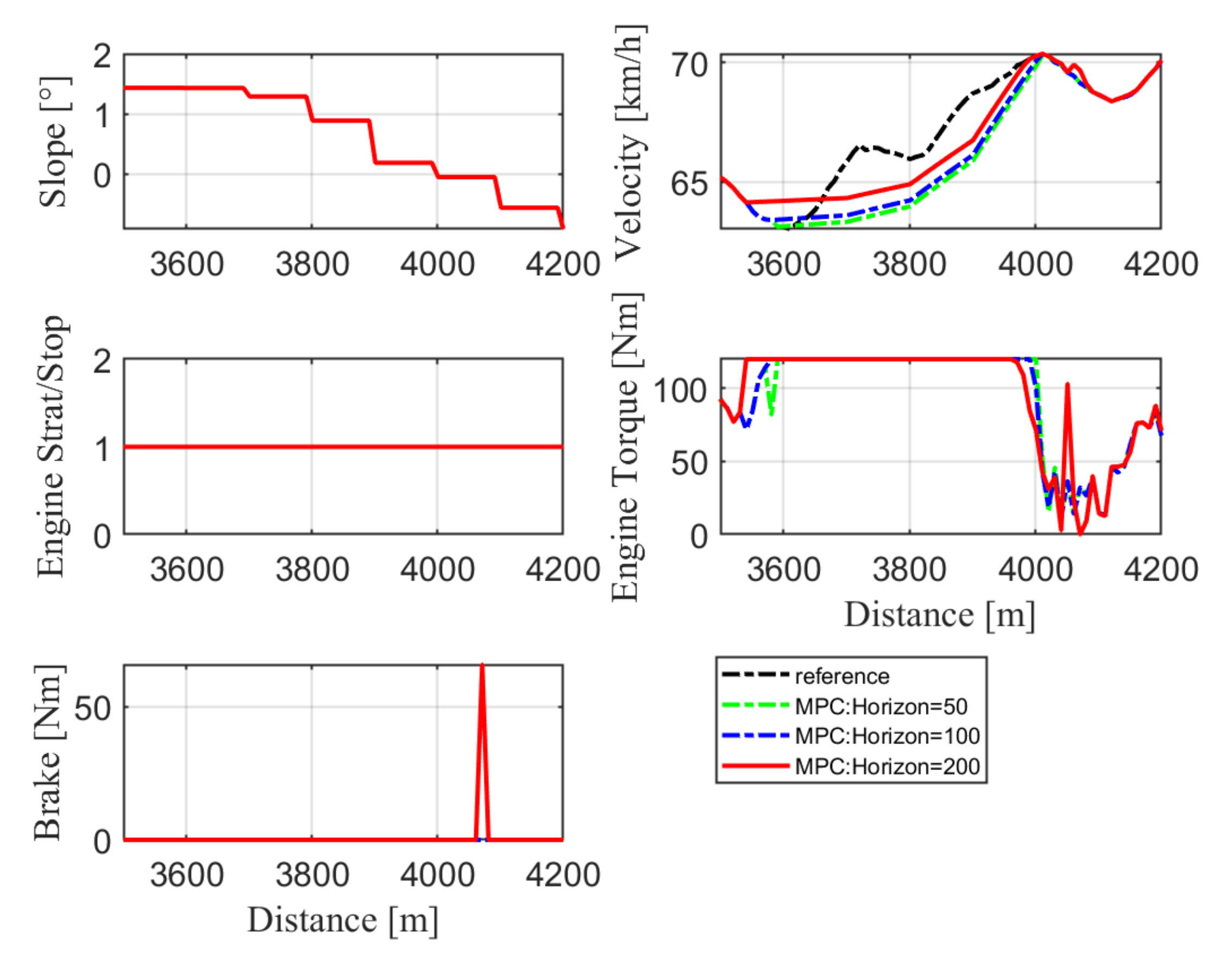}
\caption{Simulation comparison with engine start/stop method over different length of prediction horizon}
\label{Fig14}
\end{center}
\end{figure}

\begin{figure}
\begin{center}
\includegraphics[width=3.2 in]{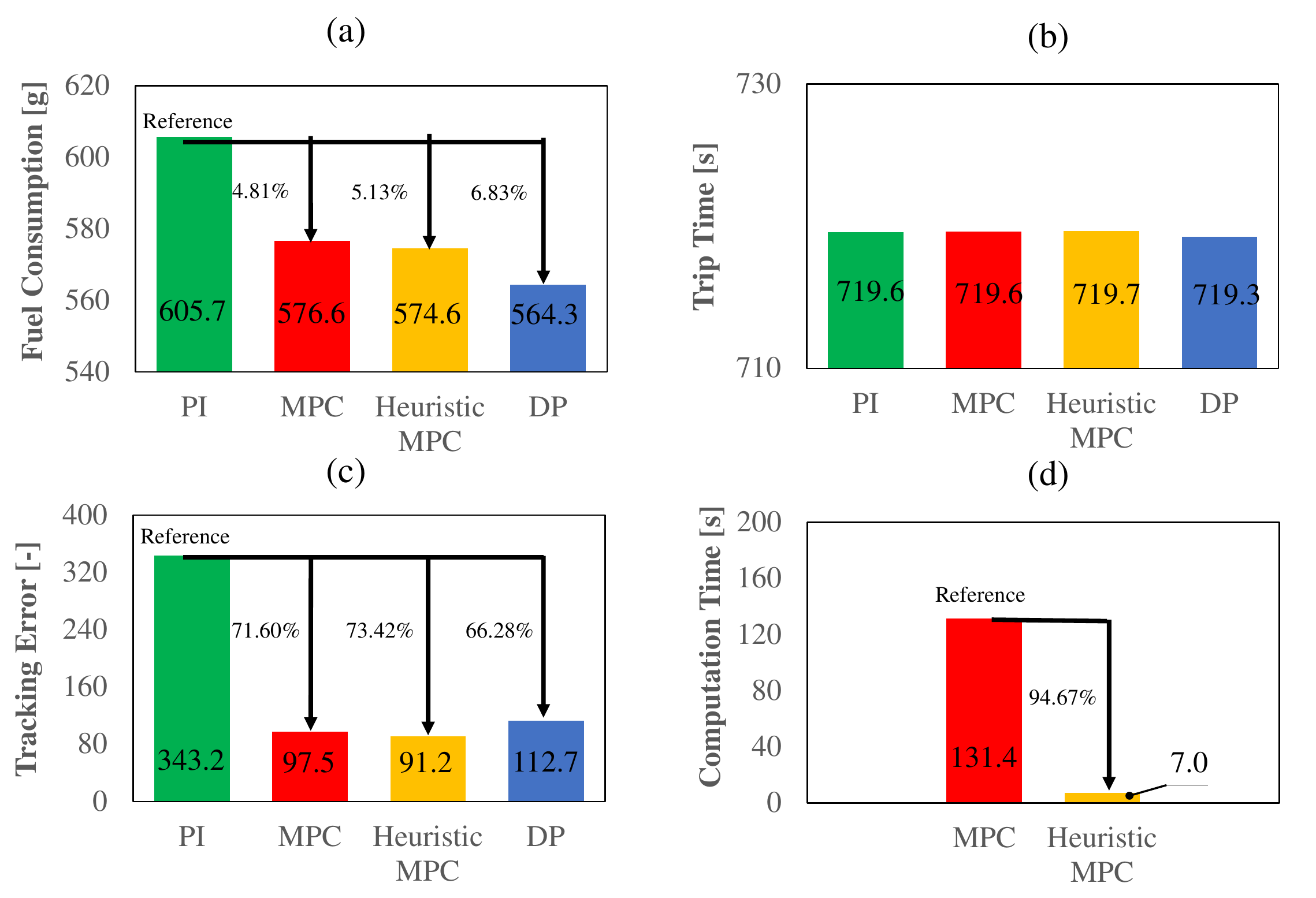}
\caption{Simulation comparison with engine start/stop method: PI vs MPC vs Heuristic MPC vs DP. (a) The fuel consumption; (b) The travel time; (c) The tracking error; (d) The computation time for obtaining our numerical solution at each instant }
\label{Fig15}
\end{center}
\end{figure}

\begin{figure}
\begin{center}
\includegraphics[width=3.2 in]{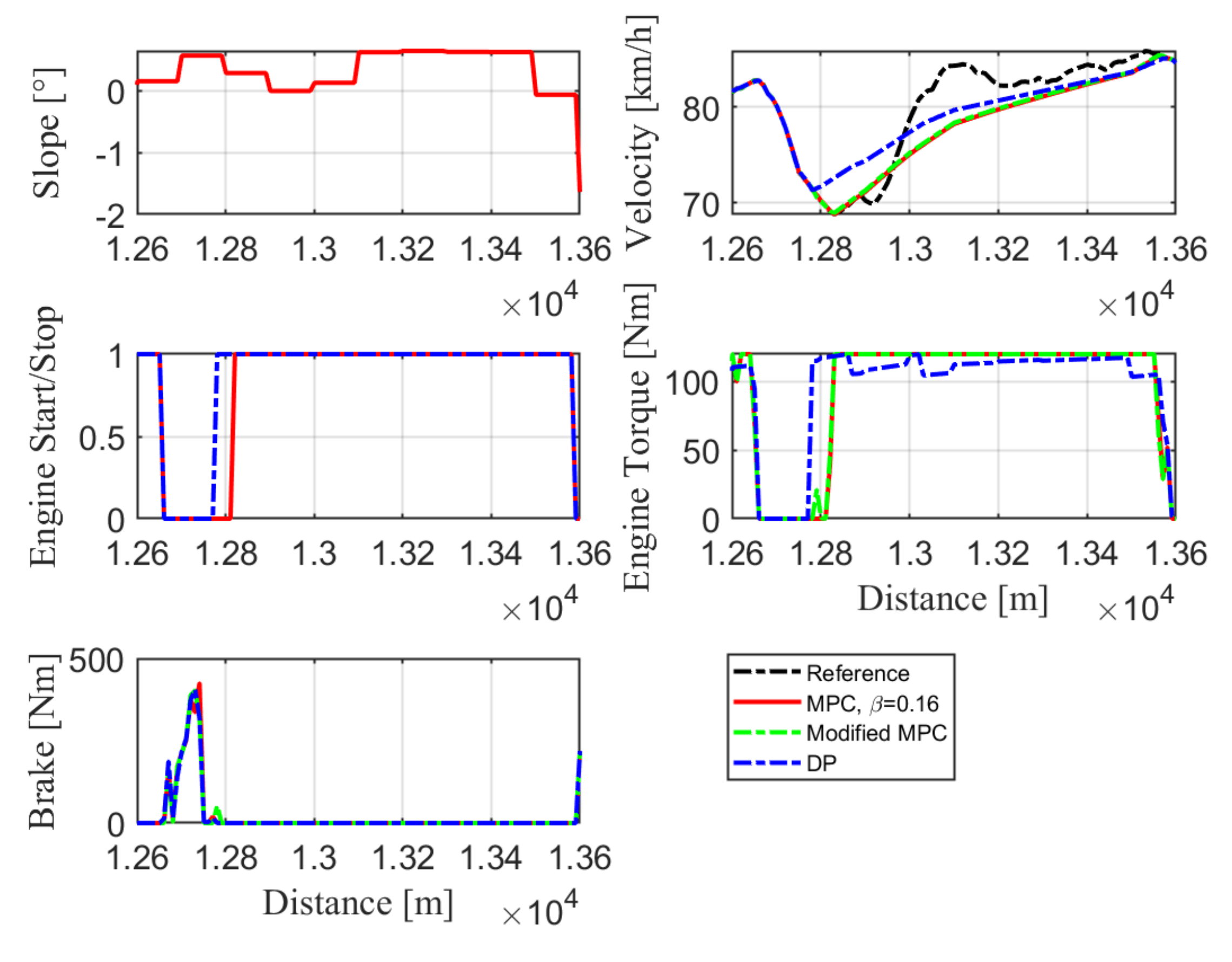}
\caption{Simulation comparison with engine start/stop method zoomed from 12600 $m$ to 13600 $m$ of the Wuhan slope profile: MPC vs Heuristic MPC vs DP }
\label{Fig16}
\end{center}
\end{figure}

\begin{figure}
\begin{center}
\includegraphics[width=3.2 in]{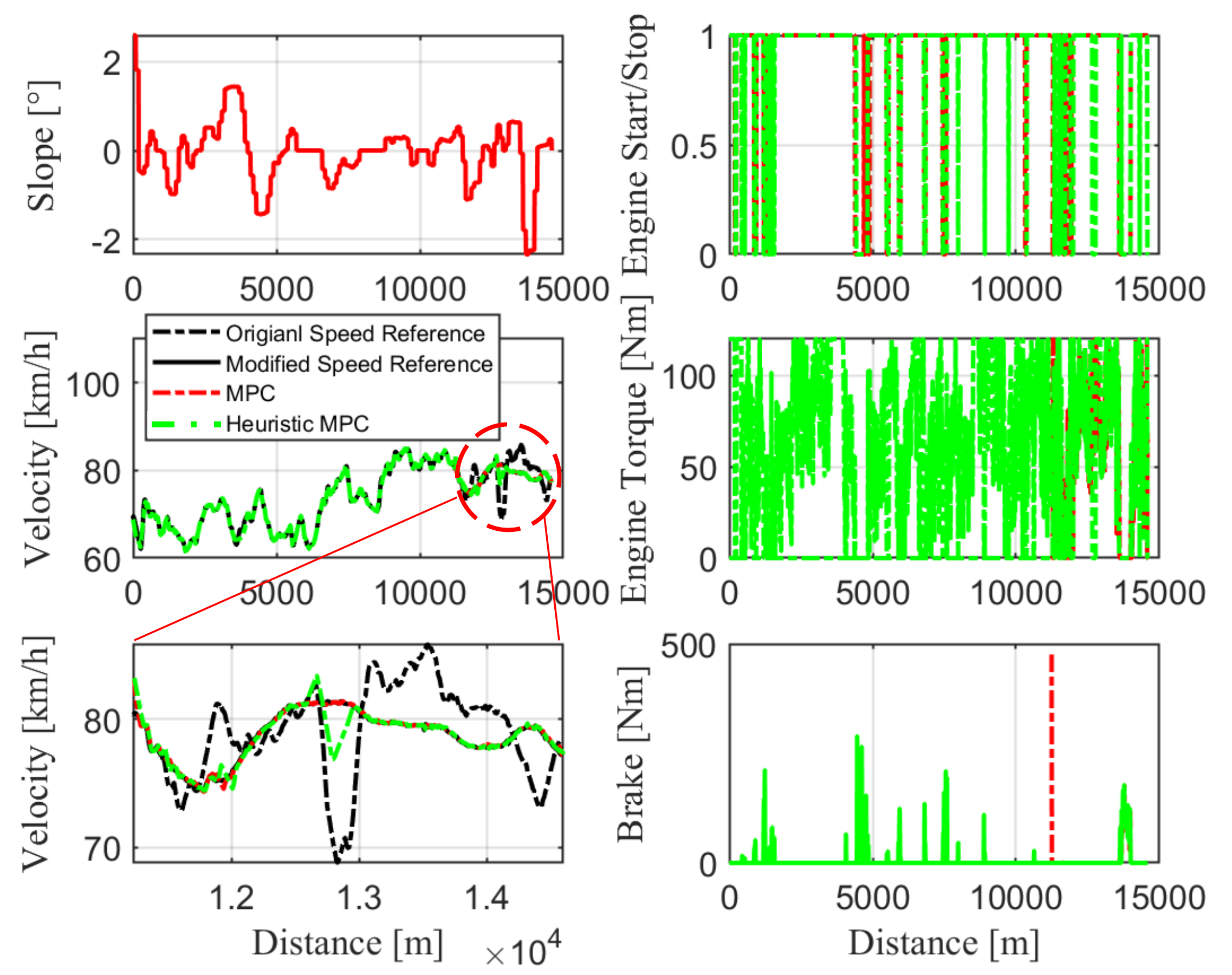}
\caption{Robustness analysis with engine start/stop method: MPC vs Heuristic MPC}
\label{Fig17}
\end{center}
\end{figure}

\begin{figure}
\begin{center}
\includegraphics[width=3.2 in]{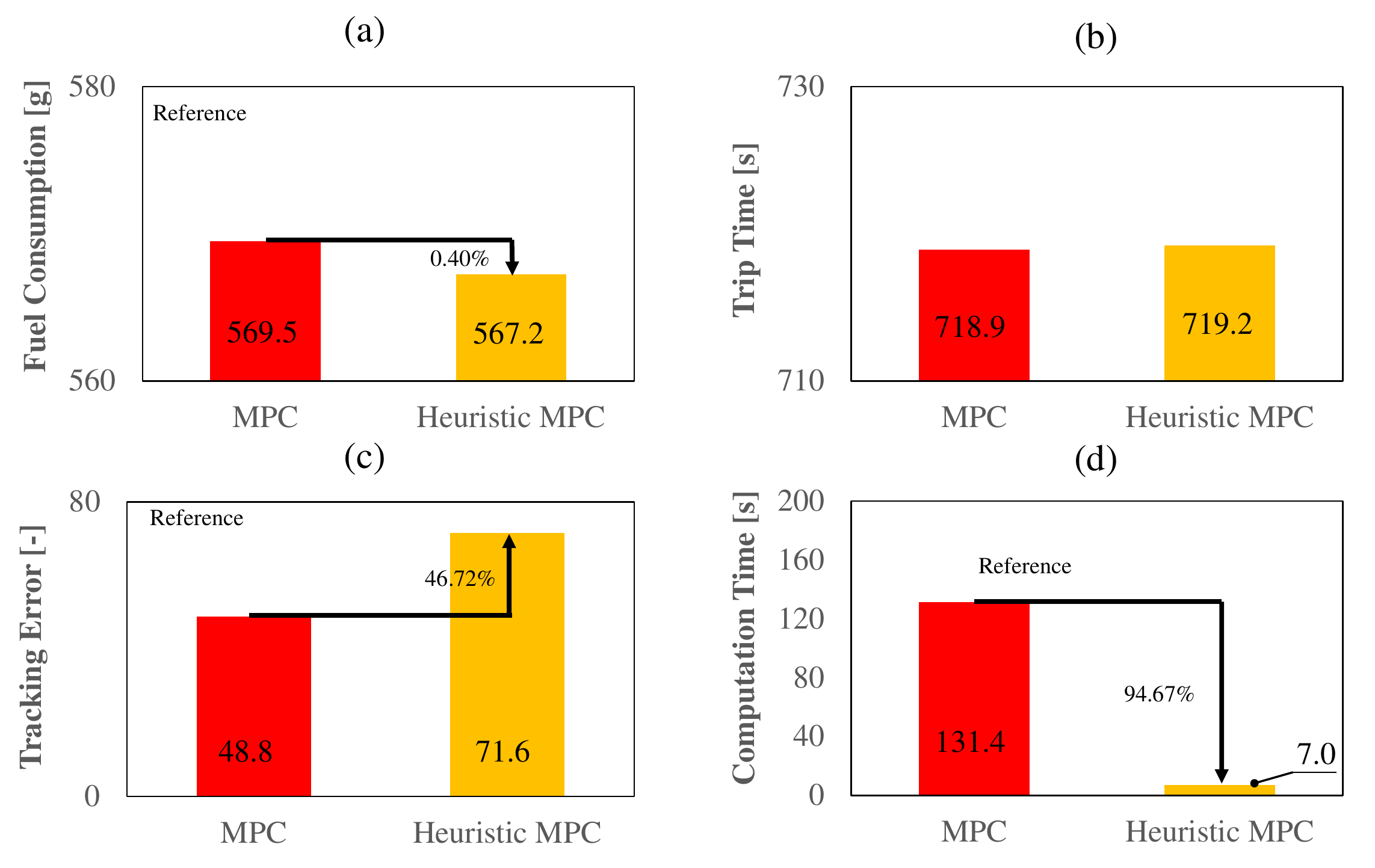}
\caption{Robustness analysis with engine start/stop method: MPC vs Heuristic MPC. (a) The fuel consumption; (b) The travel time; (c) The tracking error; (d) The computation time for obtaining our numerical solution at each instant }
\label{Fig18}
\end{center}
\end{figure}

To demonstrate the effect of the prediction horizon on MPC performance, the engine start/stop method is used as an example. The computations are performed on the MATLAB R2019a platform running on an AMD Ryzen 5 2600X 3.60-GHz with 16.0-GB RAM. Fig.~\ref{Fig13} shows that by increasing the prediction horizon from 50 m to 100 m, the tracking error will decrease from $31.89\%$ to $8.21\%$ and the fuel consumption will decrease by $1.21\%$. The reason is that increasing the prediction horizon makes more reference speed information available for the controller. MPC controller with a longer prediction horizon can accelerate earlier to track the reference speed precisely and coast earlier on the downhill to leverage the kinetic energy (Fig.~\ref{Fig14}). However, extending the prediction horizon increases the computation footprint significantly as shown in Fig.~\ref{Fig13} (d). It also makes the MPC susceptible to speed prediction error. The fuel consumption is increased by extending the horizon since the tracking error is emphasized in the cost function. The controller will calculate more engine torque for a better tracking performance with a longer prediction horizon (Fig.~\ref{Fig14}).

\subsection{Heuristic MPC}

Solving such an MINLP problem as shown in (\ref{cost of engine start/stop}) and (\ref{constraints of engine start/stop}) is computationally demanding. To eliminate the computational burden for real-time implementation, the offline DP result of the engine start/stop sequence is utilized in the MPC considering the fact that the engine start/stop decision is settled to some extent for given road grade and reference speed~\cite{biasini2013near}. Then the optimal problem is tailored as 
\begin{equation}
\begin{aligned}
\min _{T_{e}(\cdot | k), T_{b}(\cdot | k) } \sum_{i=0}^{N_{h}}(& \beta \cdot\left(M_{f}(i | k)\right)+\\
& (1-\beta) \cdot\left(v(i | k)-v_{\text{ref}} \right)^{2}) \Delta s
\end{aligned}
\end{equation}
subject to:\\
\begin{equation}
\begin{aligned}
&v(i+1|k) =v(i|k) +\frac{1}{m_{\text{eff}} v(i|k)}\left[F_{t}-F_{r}\right] \Delta s, i=0: N-1\\
& 50 km/h \leq v(i | k) \leq 90 km/h,  i=0: N \\
& 0 Nm \leq T_{e}(i | k) \leq 120 Nm,  i=0: N-1 \\
& 0 Nm \leq T_{b}(i | k) \leq 500 Nm,  i=0: N-1 \\
&v(0)=75 km/h.
\end{aligned}
\end{equation}
The integer variable $d$ and the operational constraints $\delta(\tau) \leq 1-(\delta(j-1)-\delta(j))$, which avoids the frequent switch, are dropped from the original problem definition in (\ref{cost of engine start/stop}) and (\ref{constraints of engine start/stop}). The simulation result shows that the computation time will reduce dramatically. As shown in Fig.~\ref{Fig15}(d), the computation time at each instance of the heuristic MPC is reduced to 7.0 s. Note that there is still a lot of work to do to further reduce the computation times, for instance, by replacing the commercial NLP solver with more efficient or tailored NLP solvers~\cite{liao2020fbstab}, exploiting real-time iteration strategy~\cite{henriksson2004flexible}, converting the MATLAB code to C code. These are left for future research.

The fuel consumption of this heuristic MPC is better than the MPC result due to the use of the offline engine start/stop results calculated with DP, and only slightly worse than the DP results. As shown in Fig.~\ref{Fig16}, the heuristic MPC will not even accelerate when the engine is turned on till the reference speed increased dramatically over the prediction horizon.

However, the offline calculated engine start/stop sequence depends to a larger extent on the road grade and driver’s demand (i.e., reference speed). When the reference speed is changed from 11240 m to 14600 m as shown in Fig.~\ref{Fig17}, the MPC can track the reference speed accurately. But the heuristic MPC, which uses the pre-calculated engine start/stop sequence derived from the original reference speed, shows a trajectory deviation around 12900 m due to the inappropriate engine off signal from the results of DP. The tracking error of heuristic MPC is increased by 46.72 \% since the engine start/stop sequence is not updated with the modified speed reference, while the fuel consumption and travel time are almost the same (Fig.~\ref{Fig18}). 

\section{Conclusion}

In this paper, two eco-coasting strategies are proposed using road grade preview information. The evaluation and online implementation are investigated with DP and MPC respectively. The dynamic equation considering the engine restart cost will not only give a more precise evaluation but also avoid the frequent switch of the engine start/stop signal. Finally, the online performance is evaluated by reformulating the optimal control problems of eco-coasting strategies as MIMPC problems. Simulation results show that the MIMPC improves fuel consumption to a level comparable to DP without sacrificing the travel time for both fuel cut-off and engine start/stop methods. The heuristic MPC can reduce the computation footprint by utilizing the pre-calculated engine start/stop sequence derived from DP results when the road grade and driver's demand (e.g., reference speed) are known in advance. As for the implementation of the eco-coasting strategy in the future, the V2X information can help to identify the reference speed, the pre-calculated coasting signal can be derived with cloud computing before departure. 

Since the mixed integer nonlinear programming in each discrete distance instance requires high computational cost, an algorithm for solving the MIMPC in real-time is a subject left for future research.


%



\section*{Acknowledgment}
Y. Yan acknowledges the support from the Chinese Scholarship Council that made this work possible.
This work was done when he was a visiting Ph.D. student in the Department of Naval Architecture and Marine Engineering, University of Michigan, Ann Arbor, MI 48109 USA. The work of B. Z. Gao and H. Chen was supported by the China Automobile Industry Innovation and Development Joint Fund under Grant No.U1864206, and in part by Jilin provincial Science $\&$ Technology Department No.20200301011RQ.

\ifCLASSOPTIONcaptionsoff
  \newpage
\fi



%


\bibliographystyle{IEEEtran}
\bibliography{sample}

\begin{thebibliography}{10}
\providecommand{\url}[1]{#1}
\csname url@samestyle\endcsname
\providecommand{\newblock}{\relax}
\providecommand{\bibinfo}[2]{#2}
\providecommand{\BIBentrySTDinterwordspacing}{\spaceskip=0pt\relax}
\providecommand{\BIBentryALTinterwordstretchfactor}{4}
\providecommand{\BIBentryALTinterwordspacing}{\spaceskip=\fontdimen2\font plus
\BIBentryALTinterwordstretchfactor\fontdimen3\font minus
  \fontdimen4\font\relax}
\providecommand{\BIBforeignlanguage}[2]{{%
\expandafter\ifx\csname l@#1\endcsname\relax
\typeout{** WARNING: IEEEtran.bst: No hyphenation pattern has been}%
\typeout{** loaded for the language `#1'. Using the pattern for}%
\typeout{** the default language instead.}%
\else
\language=\csname l@#1\endcsname
\fi
#2}}
\providecommand{\BIBdecl}{\relax}
\BIBdecl

\bibitem{varnhagen2017electronic}
R.~Varnhagen, ``Electronic horizon: A map as a sensor and predictive control,''
  SAE Technical Paper, Tech. Rep., 2017.

\bibitem{salgueiredo2017experimental}
C.~F. Salgueiredo, O.~Orfila, G.~Saint~Pierre, P.~Doublet, S.~Glaser,
  S.~Doncieux, and V.~Billat, ``Experimental testing and simulations of speed
  variations impact on fuel consumption of conventional gasoline passenger
  cars,'' \emph{Transportation Research Part D: Transport and Environment},
  vol.~57, pp. 336--349, 2017.

\bibitem{li2015effect}
S.~E. Li, K.~Deng, Y.~Zheng, and H.~Peng, ``Effect of pulse-and-glide strategy
  on traffic flow for a platoon of mixed automated and manually driven
  vehicles,'' \emph{Computer-Aided Civil and Infrastructure Engineering},
  vol.~30, no.~11, pp. 892--905, 2015.

\bibitem{balluchi1997cut}
A.~Balluchi, M.~Di~Benedetto, C.~Pinello, C.~Rossi, and
  A.~Sangiovanni-Vincentelli, ``Cut-off in engine control: a hybrid system
  approach,'' in \emph{Proceedings of the 36th IEEE Conference on Decision and
  Control}, vol.~5.\hskip 1em plus 0.5em minus 0.4em\relax IEEE, 1997, pp.
  4720--4725.

\bibitem{katsumata2008development}
Y.~Katsumata, S.~Segawa, K.~Adachi, A.~Higashimata, and Y.~Ochi, ``Development
  of a slip speed control system for a lock-up clutch (part ii),'' SAE
  Technical Paper, Tech. Rep., 2008.

\bibitem{koch2014criteria}
H.~Koch-Groeber and J.~Wang, ``Criteria for coasting on highways for passenger
  cars,'' SAE Technical Paper, Tech. Rep., 2014.

\bibitem{lee2009vehicle}
J.~Lee, ``Vehicle inertia impact on fuel consumption of conventional and hybrid
  electric vehicles using acceleration and coast driving strategy,'' Ph.D.
  dissertation, Virginia Tech, 2009.

\bibitem{bishop2007engine}
J.~Bishop, A.~Nedungadi, G.~Ostrowski, B.~Surampudi, P.~Armiroli, and
  E.~Taspinar, ``An engine start/stop system for improved fuel economy,'' SAE
  technical paper, Tech. Rep., 2007.

\bibitem{griefnow2017next}
P.~Griefnow, J.~Andert, and D.~Jolovic, ``Next-generation low-voltage power
  nets impacts of advanced stop/start and sailing functionalities,'' \emph{SAE
  International Journal of Fuels and Lubricants}, vol.~10, no.~2, pp. 556--573,
  2017.

\bibitem{griefnow2019real}
P.~Griefnow, J.~Andert, F.~Xia, S.~Klein, P.~Stoffel, M.~Engels, and
  D.~Jolovic, ``Real-time modeling of a 48v p0 mild hybrid vehicle with
  electric compressor for model predictive control,'' \emph{SAE Tech. Paper},
  2019.

\bibitem{sohn2020analysis}
C.~Sohn, J.~Andert, and D.~Jolovic, ``An analysis of the tradeoff between fuel
  consumption and ride comfort for the pulse and glide driving strategy,''
  \emph{IEEE Transactions on Vehicular Technology}, vol.~69, no.~7, pp.
  7223--7233, 2020.

\bibitem{choi2013optimal}
S.~Choi, K.~Ko, and I.~Jeung, ``Optimal fuel-cut driving method for better fuel
  economy,'' \emph{International Journal of Automotive Technology}, vol.~14,
  no.~2, pp. 183--187, 2013.

\bibitem{mueller2011next}
N.~Mueller, S.~Strauss, S.~Tumback, G.-C. Goh, and A.~Christ, ``Next generation
  engine start/stop systems: “free-wheeling”,'' \emph{SAE International
  Journal of Engines}, vol.~4, no.~1, pp. 874--887, 2011.

\bibitem{hellstrom2010design}
E.~Hellstr{\"o}m, J.~{\AA}slund, and L.~Nielsen, ``Design of an efficient
  algorithm for fuel-optimal look-ahead control,'' \emph{Control Engineering
  Practice}, vol.~18, no.~11, pp. 1318--1327, 2010.

\bibitem{ngo2012gear}
D.~V. Ngo, ``Gear shift strategies for automotive transmissions,'' \emph{Ph.D.
  Thesis}, 2012.

\bibitem{li2017hybrid}
G.~Li and D.~Goerges, ``Hybrid modeling and predictive control of the power
  split and gear shift in hybrid electric vehicles,'' in \emph{2017 IEEE
  Vehicle Power and Propulsion Conference (VPPC)}.\hskip 1em plus 0.5em minus
  0.4em\relax IEEE, 2017, pp. 1--6.

\bibitem{yan2012hybrid}
F.~Yan, J.~Wang, and K.~Huang, ``Hybrid electric vehicle model predictive
  control torque-split strategy incorporating engine transient
  characteristics,'' \emph{IEEE transactions on vehicular technology}, vol.~61,
  no.~6, pp. 2458--2467, 2012.

\bibitem{esmaeili2018optimal}
S.~Esmaeili, S.~Jadid, A.~Anvari-Moghaddam, and J.~M. Guerrero, ``Optimal
  operational scheduling of smart microgrids considering hourly
  reconfiguration,'' in \emph{2018 IEEE 4th Southern Power Electronics
  Conference (SPEC)}.\hskip 1em plus 0.5em minus 0.4em\relax IEEE, 2018, pp.
  1--6.

\bibitem{parisio2016stochastic}
A.~Parisio, E.~Rikos, and L.~Glielmo, ``Stochastic model predictive control for
  economic/environmental operation management of microgrids: An experimental
  case study,'' \emph{Journal of Process Control}, vol.~43, pp. 24--37, 2016.

\bibitem{mayer2015management}
B.~Mayer, M.~Killian, and M.~Kozek, ``Management of hybrid energy supply
  systems in buildings using mixed-integer model predictive control,''
  \emph{Energy conversion and management}, vol.~98, pp. 470--483, 2015.

\bibitem{belotti2013mixed}
P.~Belotti, C.~Kirches, S.~Leyffer, J.~Linderoth, J.~Luedtke, and A.~Mahajan,
  ``Mixed-integer nonlinear optimization,'' \emph{Acta Numerica}, vol.~22, pp.
  1--131, 2013.

\bibitem{takapoui2020simple}
R.~Takapoui, N.~Moehle, S.~Boyd, and A.~Bemporad, ``A simple effective
  heuristic for embedded mixed-integer quadratic programming,''
  \emph{International journal of control}, vol.~93, no.~1, pp. 2--12, 2020.

\bibitem{yu2020mixed}
H.~Yu, F.~Zhang, J.~Xi, and D.~Cao, ``Mixed-integer optimal design and energy
  management of hybrid electric vehicles with automated manual transmissions,''
  \emph{IEEE Transactions on Vehicular Technology}, vol.~69, no.~11, pp.
  12\,705--12\,715, 2020.

\bibitem{deng2020flexible}
J.~Deng, F.~Meier, and L.~del Re, ``Flexible predictive hybrid powertrain
  management with v2x information,'' in \emph{2020 59th IEEE Conference on
  Decision and Control (CDC)}.\hskip 1em plus 0.5em minus 0.4em\relax IEEE,
  2020, pp. 3500--3505.

\bibitem{elbert2014engine}
P.~Elbert, T.~N{\"u}esch, A.~Ritter, N.~Murgovski, and L.~Guzzella, ``Engine
  on/off control for the energy management of a serial hybrid electric bus via
  convex optimization,'' \emph{IEEE Transactions on Vehicular Technology},
  vol.~63, no.~8, pp. 3549--3559, 2014.

\bibitem{hadj2016convex}
S.~Hadj-Said, G.~Colin, A.~Ketfi-Cherif, and Y.~Chamaillard, ``Convex
  optimization for energy management of parallel hybrid electric vehicles,''
  \emph{IFAC-PapersOnLine}, vol.~49, no.~11, pp. 271--276, 2016.

\bibitem{zhang2021hierarchical}
H.~Zhang, Q.~Fan, S.~Liu, S.~E. Li, J.~Huang, and Z.~Wang, ``Hierarchical
  energy management strategy for plug-in hybrid electric powertrain integrated
  with dual-mode combustion engine,'' \emph{Applied Energy}, vol. 304, p.
  117869, 2021.

\bibitem{han2019fundamentals}
J.~Han, A.~Vahidi, and A.~Sciarretta, ``Fundamentals of energy efficient
  driving for combustion engine and electric vehicles: An optimal control
  perspective,'' \emph{Automatica}, vol. 103, pp. 558--572, 2019.

\bibitem{jia2020energy}
Y.~Jia and D.~G{\"o}rges, ``Energy-optimal adaptive cruise control based on
  hybrid model predictive control with mixed-integer quadratic programming,''
  in \emph{2020 European Control Conference (ECC)}.\hskip 1em plus 0.5em minus
  0.4em\relax IEEE, 2020, pp. 686--692.

\bibitem{sadek2016fpga}
U.~Sadek, A.~Sarja{\v{s}}, A.~Chowdhury, and R.~Sve{\v{c}}ko, ``Fpga-based
  optimal robust minimal-order controller structure of a dc--dc converter with
  pareto front solution,'' \emph{Control Engineering Practice}, vol.~55, pp.
  149--161, 2016.

\bibitem{chu2018predictive}
H.~Chu, L.~Guo, B.~Gao, H.~Chen, N.~Bian, and J.~Zhou, ``Predictive cruise
  control using high-definition map and real vehicle implementation,''
  \emph{IEEE Transactions on Vehicular Technology}, vol.~67, no.~12, pp.
  11\,377--11\,389, 2018.

\bibitem{hou2014approximate}
C.~Hou, M.~Ouyang, L.~Xu, and H.~Wang, ``Approximate pontryagin’s minimum
  principle applied to the energy management of plug-in hybrid electric
  vehicles,'' \emph{Applied Energy}, vol. 115, pp. 174--189, 2014.

\bibitem{li2009path}
Z.~Li, J.~Sun, and S.~Oh, ``Path following for marine surface vessels with
  rudder and roll constraints: An mpc approach,'' in \emph{2009 American
  control conference}.\hskip 1em plus 0.5em minus 0.4em\relax IEEE, 2009, pp.
  3611--3616.

\bibitem{carrion2006computationally}
M.~Carri{\'o}n and J.~M. Arroyo, ``A computationally efficient mixed-integer
  linear formulation for the thermal unit commitment problem,'' \emph{IEEE
  Transactions on power systems}, vol.~21, no.~3, pp. 1371--1378, 2006.

\bibitem{lofberg2004yalmip}
J.~Lofberg, ``Yalmip: A toolbox for modeling and optimization in matlab,'' in
  \emph{2004 IEEE international conference on robotics and automation (IEEE
  Cat. No. 04CH37508)}.\hskip 1em plus 0.5em minus 0.4em\relax IEEE, 2004, pp.
  284--289.

\bibitem{tawarmalani2005polyhedral}
M.~Tawarmalani and N.~V. Sahinidis, ``A polyhedral branch-and-cut approach to
  global optimization,'' \emph{Mathematical programming}, vol. 103, no.~2, pp.
  225--249, 2005.

\bibitem{biasini2013near}
R.~Biasini, S.~Onori, and G.~Rizzoni, ``A near-optimal rule-based energy
  management strategy for medium duty hybrid truck,'' \emph{International
  Journal of Powertrains}, vol.~2, no. 2-3, pp. 232--261, 2013.

\bibitem{liao2020fbstab}
D.~Liao-McPherson and I.~Kolmanovsky, ``Fbstab: A proximally stabilized
  semismooth algorithm for convex quadratic programming,'' \emph{Automatica},
  vol. 113, p. 108801, 2020.

\bibitem{henriksson2004flexible}
D.~Henriksson and J.~{\AA}kesson, ``Flexible implementation of model predictive
  control using sub-optimal solutions,'' \emph{Department of Automatic Control,
  Lund Institute of Technology (LTH)}, 2004.

\end{thebibliography}

%








\end{document}